\documentclass[notitlepage]{jfm}

\usepackage{graphicx}
\usepackage{newtxtext}
\usepackage{newtxmath}
\usepackage{natbib}
\usepackage{hyperref}
\hypersetup{
    colorlinks = true,
    urlcolor   = blue,
    citecolor  = blue,
}

\newcommand{\RomanNumeralCaps}[1]
\linenumbers
\usepackage{csquotes}
\DeclareMathOperator{\arcsinh}{sinh^{-1}}
\usepackage{comment}
\usepackage{xcolor}

\title{Onset of thermo-convective instabilities in two-layer binary fluid systems}

\author{Saumyakanta Mishra,
\and S. V. Diwakar
    \corresp{\email{diwakar@jncasr.ac.in}}}

\affiliation{Engineering Mechanics Unit, Jawaharlal Nehru Centre for Advanced Scientific Research, Jakkur, Bengaluru 560064, India}

\begin{document}

\maketitle

\begin{abstract}
The current work analyses the onset characteristics of buoyancy and thermocapillary-driven instabilities in two-layer binary fluid systems near their upper critical solution temperature (UCST). Owing to the non-trivial thickness of fluid interfaces in such regimes, the present analysis utilises the phase-field approach with a modified free energy expression to capture the temperature-dependent solubility of the fluids and their interfacial width. In addition, a spectral collocation-based discretisation has been incorporated along with a suitable grid mapping strategy. Consequently, the neutral curves corresponding to the flow onset are accurately predicted for different fluid combinations. In the case of pure buoyancy-driven (Rayleigh-B{\'e}nard) convection, the parametric range for oscillatory onset is found to shrink when the system approaches UCST, as the increased solubility results in less favourable conditions for oscillatory onset. The marginal stability curves of each fluid combination exhibit their own drift pattern based on the thermo-physical and transport properties. For systems with added thermocapillarity effects (Rayleigh-B{\'e}nard-Marangoni convection), the changing solubilities and the interfacial thickness, like the interfacial tension, exhibit a dual role that results in system-specific expansion/shrinkage of the parametric space for oscillatory flow onset.

\end{abstract}


\section{Introduction}

The phenomenon of Rayleigh-B{\'e}nard-Marangoni (RBM) convection in multiple fluid layers has attracted broader interests due to its facets of pattern formation \citep{Newell_Whitehead_1969} and its relevance in diverse scenarios such as Earth mantle convection \citep{Richter1974, Busse1981} and liquid encapsulated crystal growth \citep{Johnson1975, Shen1990}. The multi-modal interactions between the layers in such systems manifest varied and complex patterns that are typically classified based on the number of layers that undergo primary excitation. When the convection is dominant in only one of the layers, it is commonly referred to as the \enquote{dragging mode} \citep{Johnson1997}, wherein the other passive layers are driven by the continuity of velocity and shear stress at the fluid interfaces. When more than one layer undergoes primary excitation, two distinct non-oscillatory modes can occur: the mechanical coupling (counter-rotating rolls) mode and the thermal coupling (co-rotating rolls) mode. For specific fluid properties and system configurations, one can also observe oscillatory patterns involving cyclic variation of the system between the above non-oscillatory states \citep{Rasenat1989}. Typically, such oscillatory onset occurs when there is competition between a minimum of two modes, i.e., between a bulk mode and an interfacial mode or between two bulk modes. Linear stability analysis of a simple two-layer RB problem \citep{Renardy1985} reveals that the system can be non-self-adjoint and exhibit Hopf bifurcation for fluids with certain favourable property combinations. \citet{Renardy1996} showed that for infinite Prandtl number and pure buoyancy-driven convection, the system would be oscillatory when $\rho\beta\alpha \gg 1.0$ or $\rho\beta\alpha \ll 1.0$ ($\rho$: ratio of densities, $\beta$: ratio of coefficients of thermal expansion, $\alpha$: ratio of thermal diffusivities). \citet{Degen1998} provided experimental confirmation of these findings by observing time-dependent onset in the water - 47v2 silicone oil combination with a \(\rho\beta\alpha\) value of 0.375. Using the concept of balanced contrasts \citep{Colinet1994}, \citet{Diwakar2014JFM} showed that in addition to favourable \(\rho\beta\alpha\), the occurrence of oscillatory modes also depends on the critical height ratio ($a^*$) of the system, at which the Rayleigh numbers of the two layers are equal.

An essential consideration in all the above works is the premise of perfect immiscibility between the fluids, i.e., zero interfacial thickness. However, such a representation only offers a partial description of fluid systems and precludes the near-critical (consolute) behaviour of fluids wherein their interface is a region of finite thickness \citep{Rayleigh1892,vanderWalls1893} with a smooth but rapid property change. It is well known from mixture thermodynamics that a tiny amount of miscibility exists between the so-called immiscible fluids, even at low temperatures \citep{Lowengrub1998}. As the temperature increases, the fluids become more soluble in one another, and their interface thickens. Eventually, they become completely miscible after a critical value, known as the upper consolute or upper critical solution temperature (UCST). The resulting system forms a single-well potential curve in the free energy vs. phase-parameter diagram. Hence, a pertinent question is how the onset behaviour of RBM convection gets modified in such binary fluids when they approach UCST from an immiscible state. In particular, it would be interesting to understand the evolution of parametric space in which oscillatory modes are manifested. 

The present work aims to address the above questions via a diffuse interface approach involving the well-known Eulerian (fixed grids) `phase-field model' \citep{Jacqmin1999} wherein the whole system is represented as a single continuum. A phase evolution equation is transiently solved to determine the fluids' distribution within the domain, and the fluid interfaces are captured implicitly. The surface tension effects are estimated from the mixing free energy of the fluids. Note that understanding the stability of fluid systems using the phase-field model is a challenging task despite its `one-fluid' framework. Several works have attempted to overcome these challenges through unique means. \citet{Borcia2003} presented a phase-field formulation to study the short and long wavelengths instabilities of Marangoni convection in liquid-gas systems. \citet{Yue2004} proposed a diffuse-interface model for micro-structured complex fluids using an energy-based variational formulation. \citet{Celani2009} studied the onset of Rayleigh-Taylor instability in immiscible fluids for small Atwood numbers via the phase-field formulation. \citet{Guo2015} proposed a thermodynamically consistent model for thermo-capillary effects and discussed the migration in density-matched fluids. While the above works focus on using the phase-field method for ``immiscible" fluids, \citet{Bestehorn2021} proposed a new free energy functional that could capture the immiscibility to miscibility phase transition in the context of Faraday instability. 

In the present work, the free energy formulation of \citet{Bestehorn2021} has been adopted in conjunction with the spectral collocation procedure to understand the stability of RBM convection in binary systems. In order to achieve better accuracy, the spectral collocation approach conventionally requires the use of the Gauss-Lobatto-Chebyshev (G-L-C) grid that involves clustering nodes in the vicinity of the domain extremities. While such a configuration is conducive to modelling single-phase systems, the need for proper resolution of sharp gradients around the diffuse interface precludes its direct usage in two-layer systems. One may find a workaround by using a vast number of GLC nodes; however, this increases the computational cost and may not be feasible in most situations. Hence, to resolve the diffuse interface better and to reduce the associated computational effort, a transformation strategy \citep{Tee2006, Diwakar2015} has been utilized here that maps the G-L-C nodes in the transformed domain to cluster around the diffuse interface in the physical domain. The clustering of points around the mixing layer region is controlled presently by a scaling parameter, \(\in\). The consistency of this mapping approach has been verified here by reproducing the results of a sharp interface approach obtained using the conventional domain decomposition method (DDM).

Apart from the implementational issues associated with the spectral version of the phase-field method, challenges also arise while accounting for the density inhomogeneity in the system. When the variation in fluids' densities is small, one could adopt a procedure similar to the classical Boussinesq approximation, wherein one neglects the density variation caused by temperature except in the body-force term. Under this assumption, a solenoidal velocity field is obtained throughout the domain. However, the scenario becomes complicated for fluid systems with considerable density differences. Such systems fall under the notion of quasi-incompressible fluids that were adopted into the diffuse-interface model by \citet{Lowengrub1998}. Though the densities of the individual fluid components remain constant, the density gradient in the diffuse interfacial region gives rise to a local non-solenoidal velocity field ($\bnabla\cdot\boldsymbol{u} \neq 0$). In other words, an extensional component of velocity that is proportional to the diffusive flux of components appears in the interfacial zone \citep{JHH1996}, thus making the velocity field non-solenoidal and the chemical potential dependent on pressure \citep{Lowengrub1998, Abels2012}. This leads to an inconsistent scenario as the kinematic conditions and not the thermodynamic constraints presently determine the pressure field \citep{Anderson1998}. It thus becomes cumbersome to carry out stability analyses for a general quasi-incompressible consideration. Fortunately, an easy remedy to this issue can be found by changing the averaging process used to arrive at the mean velocity field. Note that the velocity field is conventionally represented as a mass-averaged quantity, which leads to a non-solenoidal velocity field in the diffuse region. \citet{DSS2007} adopted a different approach wherein the mean velocity field was obtained by volume-averaging the different components' velocities. Through this means, the velocity field becomes solenoidal (${\bnabla}\cdot\boldsymbol{u} = 0$) throughout the system, including the interfacial region(s). In fact, this velocity is the same as the mass-averaged velocity in the bulk part and differs only in the transition layer. \citet{Abels2012} used this idea to develop a frame-invariant phase-field model for incompressible fluids. 

In the current work, we have thus employed the volume-averaged formulation of \citet{DSS2007} to implement a spectral phase-field stability solver for understanding the influence of diffuse interface on the onset of oscillatory convection in two-layer RBM systems. Following \citet{Colinet1994} and \citet{Diwakar2014JFM}, we consider fluid systems with different values of $\rho\beta\alpha$ and $a^{*}$ and observe how the parametric window for oscillatory convection changes when the fluid system approaches the UCST from an immiscible state. The analysis is first carried out in the context of Rayleigh-B{\'e}nard (RB) convection in the system, followed by the additional consideration of Marangoni effects. This paper contains five sections, including the current introduction. In the following section, we describe the diffuse interface model and the free energy formulation of \citet{Bestehorn2021} that mimics the immiscibility to miscibility transition. We also describe the underlying assumptions utilized to simplify the stability formulation of RBM convection. In section 3, we look at the spectral collocation implementation along with the grid transformation. We also perform a few consistency checks to validate the numerical approach. The onset of RB convection is analyzed for different fluid configurations in section 4. Here, we show how the system's propensity to exhibit oscillatory convection decreases as it approaches UCST and how each fluid system comes with its own drift pattern of stability curves, essentially determined by their thermo-physical and transport properties. Finally, section 5 deals with the onset of RBM convection in binary fluids. The solubility is shown here to act in unison with the interfacial tension to make the system either self-adjoint or non-self-adjoint. Thus, it comes with its own dual role in deciding the parametric window for the onset of oscillatory convection.  

\section{The binary fluid model}

\begin{figure}
    \includegraphics[trim = {7cm 6cm 5cm 5cm},scale=0.60]{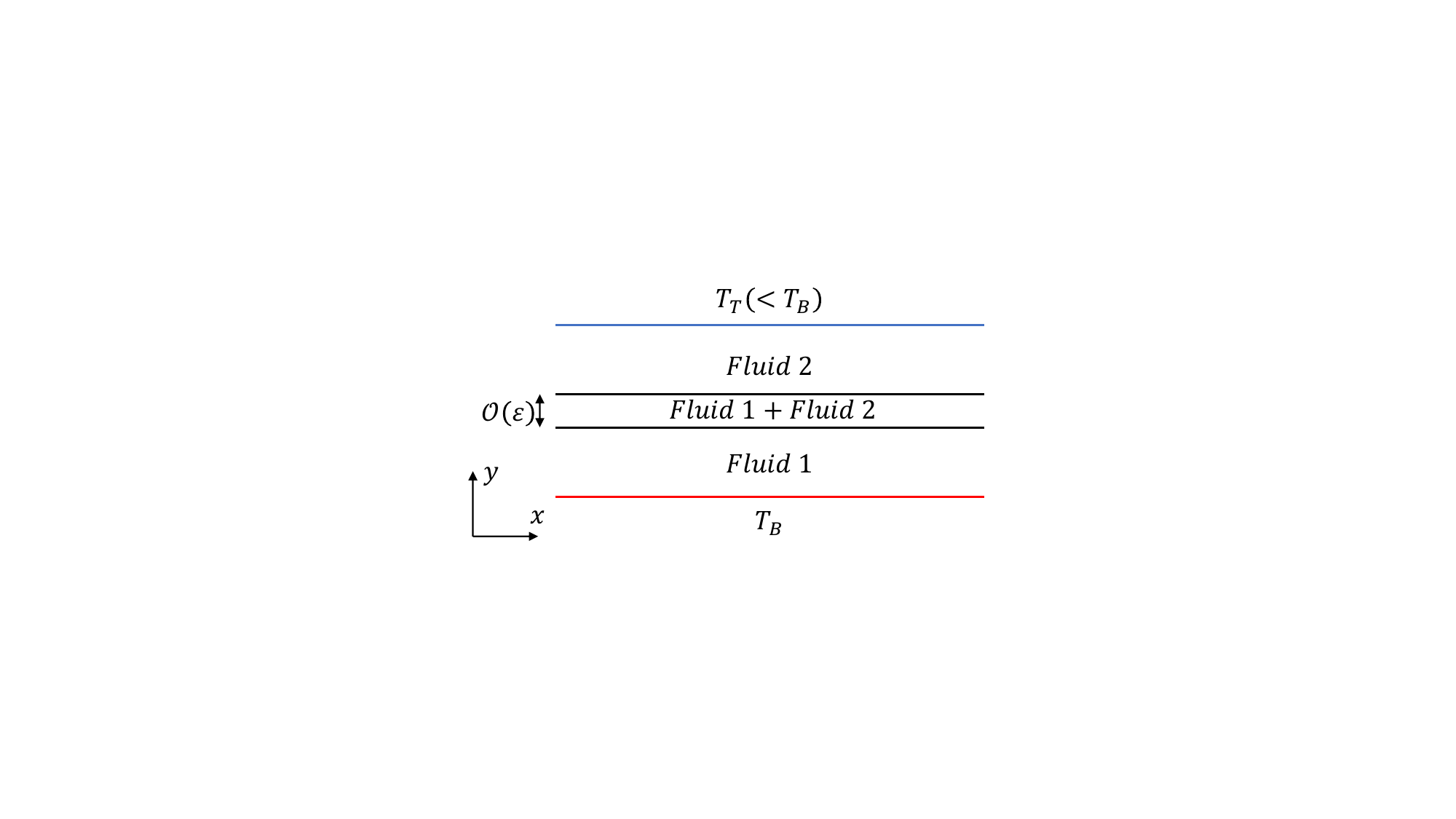}
    \caption{Schematic of R-B convection problem in binary fluids}
    \label{fig:geometry_of_problem}
\end{figure}

In this section, we define a suitable model for phase transition in binary fluids and subsequently describe the set of equations that governs the onset of RBM convection in such systems when they are slightly below UCST. The configuration of the present system is shown in Fig.~\ref{fig:geometry_of_problem} wherein the lighter fluid is stacked over the heavier one, and the system is infinite in horizontal directions. The top and bottom walls are maintained at uniform temperatures, with the bottom being hotter than the top. The fluid interface is diffuse, and its thickness diverges as the system approaches UCST. We mimic this interfacial behaviour through a phase-field model that utilizes the fluid free energy to smoothen the interfacial effects over a thin, numerically resolvable region. The simplest form of this free energy density functional \citep{Jacqmin1999} can be written as 

\begin{equation} \label{eq:basic:FE}
    f\left( \phi,\bnabla\phi \right) = \mathcal{A}_1 \Psi\left(\phi\right) + \frac{1}{2}\mathcal{A}_2\left| \bnabla\phi \right|^2, 
\end{equation}
where $\phi$ is the phase-field parameter that helps identify the bulk phases and the intervening interface. $\mathcal{A}_1$ and $\mathcal{A}_2$ are constants related to the surface tension coefficient and the thickness of the diffuse interface. The first term in the r.h.s. of the above equation represents the bulk energy part, which, for a two-phase/two-layer system, forms a double-well potential in the free energy vs phase parameter diagram. Different forms of bulk energy expression are generally used to mimic this double-well behaviour. $\Psi(\phi) = (\phi^2 - 1)^2$ \citep{Yue2004,Celani2009} is the most widely used form, where $\phi_{bp}=\pm 1$ represent the bulk phases. Other varieties like $\Psi(\phi) = (\phi^2 - 1/4)^2$ and $\Psi(\phi) = (\phi^2 - 1/4)^{3/2}$, where $\phi_{bp} = \pm 1/2$, can also be used \citep{Jacqmin1999}. The second term in Eq.~\eqref{eq:basic:FE} represents the gradient energy required for sustaining the interface between the two phases. This gradient part accounts for the weak non-local interaction between the components \citep{Yue2004} and is responsible for the formation of the mixing layer between the two fluids \citep{Celani2009}. The competition between the philic (gradient energy) and the phobic (bulk mixing energy) effects \citep{Yue2004} determines the structure of the interface. Typically, the width of the interface is proportional to $\sqrt{\mathcal{A}_2/\mathcal{A}_1}$ and the surface tension is proportional to $\sqrt{\mathcal{A}_1\mathcal{A}_2}$. In the case of miscibility transition in the binary fluids, the above double-well potential transforms to a single-well as the system temperature transitions over UCST. \citet{Bestehorn2021} proposed the following free energy functional to capture this transition.
\begin{equation} \label{eq:bestehorn}
    f\left( \phi, \bnabla\phi \right) = \mathcal{A}_1 \left[ \mathcal{H}(r) r^q \phi^4 - r \phi^2 \right] + \frac{1}{2} \mathcal{H}\left(r\right) \mathcal{A}_2 r^p \left| \bnabla\phi \right|^2,
\end{equation}
with $p$ and $q$ being the model constants. $r$ (defined below) provides information on the base operating condition of the system. \(r=1\) indicates that the system is far below the upper consolute temperature of the fluids, and its value diminishes as the system approaches the UCST. \(r\) is negative for system temperatures above UCST, i.e., when the components are completely miscible with no intervening interfaces. Thus, $r$, in combination with \(\mathcal{H}(r)\), the Heaviside step function, provides a convenient means for modelling the immiscible to the miscible phase transition. Note that \(\mathcal{H}(r)\) is zero for any negative \(r\) and is one otherwise.

Since the present R-B convection system is non-isothermal, the free energy expression in Eq.~\eqref{eq:bestehorn} needs to be modified in line with the formulation suggested by \citet{Alt1992}, and \citet{Antanovskii1995}. Correspondingly, the free energy of this non-isothermal system is given as
\begin{equation} \label{eq:general_free_energy_equation}
    F\left(\phi, \theta, \bnabla\phi \right) = \int_\Omega \left[ f_{bulk}(\phi,\theta) + \mathcal{H}(r) \frac{\Lambda}{2}  r^p \left| \bnabla \phi \right|^2 \right]  \mathrm{d\boldsymbol{x}},
\end{equation}
where
\begin{equation} \label{eq:general_bulk_free_energy_equation}
    f_{bulk}(\phi,\theta) = \rho c(\phi)\theta - \rho c(\phi)\theta \log{\theta} + \mathcal{H}(r)\frac{\Lambda}{4\epsilon^2} r^q \phi^4 - \frac{\Lambda}{2\epsilon^2} r \phi^2.
\end{equation}

Note that the first two terms in the bulk energy expression correspond to the changes in internal energy and entropy caused by changes in temperature. The above expression is similar to the classical Ginzburg-Landau free energy expression wherein \({\Lambda}\) is the magnitude of the mixing free energy of the system \citep{Celani2009}. Here, the internal energy part is considered to be a linear function of \(\phi\).

The miscibility transition parameter, $r$, is defined as
\begin{equation} \label{eq:r_define}
    r = \frac{\exp{(-a\vartheta)}-\exp{(a\vartheta)}}{\exp{(-a\vartheta)}+(Le)\exp{(a\vartheta)}} 
\end{equation}
where the reduced temperature \enquote{$\vartheta$} is defined as $(\theta-\theta_{crit})/\theta_{crit}$. $a$ is a positive constant, and \(Le\) is the Lewis number, which gives the ratio of thermal diffusivity to the mass diffusivity of the components involved. 
\begin{figure}
    \centering
    \includegraphics[scale=0.52]{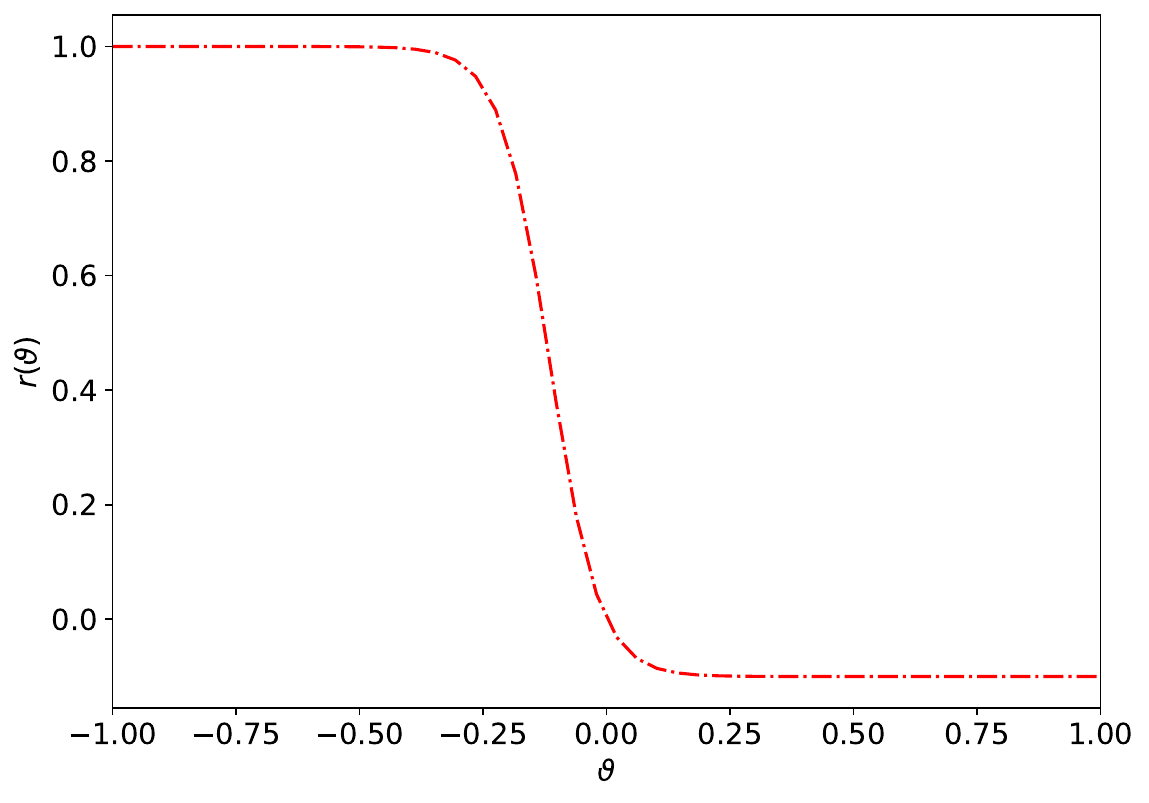}
    \caption{Variation of $r$ with $\vartheta$ for $a=10$, and $Le=10$}
    \label{fig:definition_r}
\end{figure}

Figure~\ref{fig:definition_r} gives an idea of how \(r\) varies with \(\vartheta\) for $a=10$, and $Le=10$. It can be observed that $r$ is equal to one when the system is far below the UCST, and it saturates to a value of \(-\frac{1}{Le}\) when the system is above UCST. Typically, $r$ is spatially inhomogeneous in a non-isothermal domain. However, this inhomogeneity can be deemed insignificant in the present case of two-layer RBM convection as we are only interested in the onset behaviour, which normally occurs within a critical temperature difference of the order of 1K \citep{Degen1998}. From Fig.~\ref{fig:definition_r}, it is evident that $r$ reduces from a value of one to \(-\frac{1}{Le}\) in the range of $\vartheta$ between $-0.375$ and $0.125$. In reality, this $\vartheta$ range corresponds to a large change in the absolute temperature for a binary fluid system like FC-72 and 1cSt Silicon Oil with UCST = 315.5 K. Thus, the rate of change of $r$ with temperature is meagre, which implies that any fluctuations in $r$ would be orders of magnitude lower than the temperature fluctuations. Note that the inhomogeneity in base $r$, which is proportional to the base temperature gradient, can also be further subdued by considering thicker fluid layers that will result in smaller $\Delta T_{critical}$ for the onset of buoyancy-driven convection. Hence, we consider a spatially homogeneous $r$ in the present work as it also helps make the complex problem at hand more tractable.

With the above considerations, we now formulate the governing equations for the evolution of RBM convection in the binary fluid system. Here, the phase parameter is evolved using a scalar transport equation wherein the diffusion process is driven by the gradients of chemical potential \citep{Anderson1998}, $\mu$, given as
\begin{equation} \label{eq: chemical_potential_equation}
    \mu = \frac{\delta F}{\delta\phi} = \Lambda\left[\frac{\mathcal{H}(r) r^q \phi^3 - r \phi}{\epsilon^2} -  \mathcal{H}(r)r^p \nabla^2\phi\right].
\end{equation}
We have assumed $\rho_1 c_1 \approx \rho_2 c_2$ to simplify our calculations. We mainly focus on the purely immiscible to partially miscible range where the value of \(r\) is positive, {\it, i.e.} \(0.005\leq r \leq 1.0\). This implies that \(\mathcal{H}(r) = 1\) for all further considerations. The effect of diffuse interface on the fluid momentum is modelled here via the forcing term, $\bnabla \cdot \boldsymbol{\tau}^{nv}$. The energy equation accounts for internal energy changes arising from both thermal transport and fluid mixing. Correspondingly, the coupled set of equations governing the phase evolution and mass, momentum, and energy conservation is given as 
\begin{equation}
    \frac{\partial \phi}{\partial t} + \boldsymbol{u}\cdot \bnabla\phi 
    = \bnabla \bcdot \left[\gamma \bnabla \bigg{\{}{\Lambda}\left(\frac{r^q \phi^3 - r \phi}{\epsilon^2} -  r^p \nabla^2\phi\right)\bigg{\}}  \right]
\end{equation}

\begin{equation}
     \frac{\partial \rho^* \boldsymbol{u}}{\partial t} + \bnabla\cdot(\rho^*\boldsymbol{u}\boldsymbol{u})
    = - \bnabla p + {\bnabla} \bcdot \left(\eta^* \left( \bnabla \boldsymbol{u} + (\bnabla \boldsymbol{u})^{+} - \frac{2}{3}\bnabla\bcdot\boldsymbol{u} \boldsymbol{I}\right)\right) 
    + \bnabla \bcdot \boldsymbol{\tau}^{nv}
    - \rho ^{'} g \boldsymbol{1}_y \label{eq:mom_conv}
\end{equation}
\begin{equation}
    \frac{D\rho^*}{Dt} + \rho^* \bnabla\bcdot \boldsymbol{u} = 0
\end{equation}
\begin{equation}
    \frac{\partial \rho^*(c^*\theta)}{\partial t} + \bnabla\bcdot(\rho^*\boldsymbol{u} c^*\theta) + \Lambda_0\left[\frac{r^q \phi^3 - r \phi}{\epsilon^2} - r^p \nabla^2\phi\right] \frac{D\phi}{Dt} + \mathcal{L}\bnabla\bcdot\boldsymbol{u}
    = \bnabla \bcdot (\kappa^* \bnabla \theta)
\end{equation}

where
\begin{equation}
    \boldsymbol{\tau}^{nv} = \mathcal{L} \boldsymbol{I} - \bnabla\phi \frac{\partial\mathcal{L}}{\partial(\bnabla\phi)}
    \label{eq:non-viscous_stress_equation}
\end{equation}
and \(\mathcal{L} = f_{mix}(\phi,\theta) + \mathcal{H}(r) \frac{\Lambda}{2}  r^p |\bnabla\phi(x)|^2\) is the Lagrangian energy density. $\rho^{'}$ in Eq.\eqref{eq:mom_conv} corresponds to $\rho^* - {\rho}_{0}$. ${\rho}_{0}$ is the base density at any location, which is given as $\rho_0 = \rho(\phi) = \rho _1(\frac{1+\phi}{2}) + \rho _2(\frac{1-\phi}{2})$. $\rho_1$ and $\rho_2$ are the individual (unmixed) densities of the fluids 1 and 2, respectively. Note that the present formulation precludes the direct implementation of the classical Boussinesq approximation, which assumes homogeneity of density everywhere, except in the buoyancy term that drives the convection \citep{Spiegel1960,Drazin&Reid}. Such an approximation would have been admissible only in fluid systems with very small density disparity. Since we do not impose such restrictions on the fluids, we proceed with the inhomogeneous base density formulation, as shown above. 

Incidentally, the current configuration of the system falls under the class of quasi-incompressible fluids, wherein the velocity field in the mixing region is non-solenoidal despite the bulk fluids being incompressible. This results in the chemical potential becoming a function of pressure. Since the pressure is determined here from kinematics and not from thermodynamics, the overall mathematical formulation and its analysis become complicated. It can be shown that the issue of non-solenoidal velocity in the mixing region arises from the way the velocity field has been formulated in the multi-phase domain. In the following subsection, we look at an alternate means of defining the velocity field such that it becomes solenoidal everywhere.   

\subsection{Volume-averaged velocity field}

The velocity field often considered in general `one-fluid' frameworks is inherently a mass-averaged/barycentric quantity \citep{JHH1996} given as 
\begin{equation}
    \boldsymbol{u} = \frac{\Tilde{\rho}_1\boldsymbol{u}_1 + \Tilde{\rho}_2\boldsymbol{u}_2}{\rho_0},
\end{equation}
where $\Tilde{\rho}_1$ and $\Tilde{\rho}_2$ are the apparent densities of the fluids 1 and 2, defined as $\Tilde{\rho}_1 = \rho _1(\frac{1+\phi}{2})$ and $\Tilde{\rho}_2 = \rho _2(\frac{1-\phi}{2})$. This mass-averaged velocity satisfies the classical continuity equation $\partial_t \rho_0 + {\bnabla}\cdot(\rho_0\boldsymbol{u}) = 0$. However, it results in a non-solenoidal velocity field, primarily in the diffuse interfacial region. In order to overcome this issue, we will now redefine the velocity field as a volume-averaged quantity \citep{Boyer2002, DSS2007, Abels2012} such that it becomes solenoidal throughout the system, including the interfacial region. Correspondingly, we use 
\begin{equation}
    \boldsymbol{u} = \frac{\Tilde{\rho}_1}{\rho_1} \boldsymbol{u}_1 + \frac{\Tilde{\rho}_2}{\rho_2} \boldsymbol{u}_2
\end{equation}
The above form of velocity field satisfies the mass conservation equations of the individual species, \(\partial_t \Tilde{\rho}_j + {\bnabla}\cdot(\Tilde{\rho}_j\boldsymbol{u}_j) = 0\), and also yields a solenoidal velocity field, as shown below.

\begin{equation}
\begin{split}
    {\bnabla}\cdot\boldsymbol{u} &= {\bnabla}\cdot\left( \frac{\Tilde{\rho}_1 \boldsymbol{u}_1}{\rho_1} \right) + {\bnabla}\cdot\left( \frac{\Tilde{\rho}_2 \boldsymbol{u}_2}{\rho_2} \right) \\
    &= - \partial_t \left( \frac{\Tilde{\rho}_1}{\rho_1} \right) - \partial_t \left( \frac{\Tilde{\rho}_2}{\rho_2} \right) \\
    &= - \partial_t \left( \frac{\Tilde{\rho}_1}{\rho_1} + \frac{\Tilde{\rho}_2}{\rho_2} \right) \\
    &= \partial_t (1) = 0
\end{split}
\end{equation}

Consequently, the overall mass balance equation gets modified as \citep{DSS2007}
\begin{equation}
    \partial_t \rho_0 + {\bnabla}\cdot(\rho_0\boldsymbol{u}) - \frac{\partial \rho_0}{\partial \phi}{\bnabla} \cdot \left(\gamma\bnabla\mu\right) = 0.
\end{equation}

\subsection{Governing equations and boundary conditions}
Following the above modification in the velocity averaging process, the equations governing RBM convection in binary fluid systems can be simplified by re-writing them in the non-conservation form. To account for the Marangoni effect, the magnitude of mixing free energy is considered as a linear function of temperature. Since the temperature fluctuations are expected to be small at onset, one can use the Taylor series expansion to write the free energy magnitude as $\Lambda = \Lambda_0 +\left(\partial\Lambda / \partial\theta\right) \theta$. Correspondingly, the full set of governing equations is written as 
\begin{equation}
     \frac{\partial \phi}{\partial t} + \boldsymbol{u} \cdot \bnabla\phi 
    = \bnabla \cdot \left[\gamma {\bnabla} \bigg{\{}{\Lambda}\left(\frac{r^q \phi^3 - r \phi}{\epsilon^2} -  r^p \nabla^2\phi \right)\bigg{\}}  \right] \label{feq_phi}
\end{equation}

\begin{equation}
    \bnabla \bcdot \boldsymbol{u} = 0, \label{feq_conti}
\end{equation}

\begin{eqnarray} \nonumber
     \rho_0 \left( \frac{\partial\boldsymbol{u}}{\partial t} + \boldsymbol{u}\cdot{\bnabla}\boldsymbol{u} \right)
     = & - {\bnabla} p + {\bnabla} \cdot \bigg(\eta^* \left( {\bnabla}\boldsymbol{u} + ({\bnabla}\boldsymbol{u})^{+}\right)\bigg) - \rho^{'} g \Hat{\boldsymbol{1}}_y  \qquad\qquad\quad\\
     &- r^p \Lambda \nabla^2\phi \bnabla\phi + \frac{1}{3}  r^p \left|\bnabla\phi\right|^2 \frac{\partial\Lambda}{\partial\theta} {\bnabla}\theta \nonumber \\
     &- \frac{1}{6} \Lambda r^p \bnabla \left|\bnabla\phi\right|^2 - r^p {\bnabla}\phi \frac{\partial\Lambda}{\partial\theta} {\bnabla}\theta\cdot{\bnabla}\phi, \label{feq_mom}
\end{eqnarray}

\begin{equation}
     \rho_0 c^* \bigg(\frac{\partial\theta}{\partial t} + \boldsymbol{u}\cdot{\bnabla}\theta \bigg) + \Lambda_0\left[\frac{r^q \phi^3 - r \phi}{\epsilon^2} - r^p \nabla^2\phi\right] \frac{D\phi}{Dt}
    = {\bnabla} \cdot (\kappa^*\bnabla\theta), \label{feq_ener}
\end{equation}

where $\rho ^{'} = \rho^*(\phi,\theta) - \rho_0$, and $\rho^*(\phi,\theta) = \rho _1( 1-\beta _1(\theta-\theta _0) )(\frac{1+\phi}{2}) + \rho _2( 1-\beta _2(\theta-\theta _0) )(\frac{1-\phi}{2})$. 
In Eq.\eqref{feq_mom}, note that the isotropic contributions to the reversible stress \eqref{eq:non-viscous_stress_equation} has been absorbed into the pressure term. Only the anisotropic part that contributes towards the dynamics \citep{Cates2018} has been written explicitly. 

At the top and bottom (Fig.~\ref{fig:geometry_of_problem}), no-slip walls are considered with constant temperatures. Correspondingly, we can write the boundary conditions as
\begin{equation}
    \boldsymbol{u} = 0 \,\,\,@ y = 0, H
\end{equation}
\begin{equation}
     \theta = \theta_B \,\,\,@ y = 0\\
\end{equation}
\begin{equation}
     \theta = \theta_T \,\,\,@ y = H\\   
\end{equation}

\noindent Note that the phase evolution equation obeys the no flux boundary condition owing to the consideration of impermeable no-slip walls at the top and bottom. Thus, we have

\begin{equation}
    \boldsymbol{1}_y \bcdot \bnabla\phi = 0  \,\,\,@ y = 0, H    
\end{equation}
\begin{equation}
    \boldsymbol{1}_y \bcdot \bnabla\mu = 0  \,\,\,@ y = 0, H    
\end{equation}


\noindent Since, \(\mu = \Lambda\left[\frac{r^q \phi^3 - r \phi}{\epsilon^2} - r^p \nabla^2\phi\right]\), we simplify the above conditions as 
\begin{equation}
     \frac{\partial\phi}{\partial y} =  \frac{\partial ^3\phi}{\partial y^3} = 0 \,\,\,@ y = 0, H
\end{equation}

\noindent Here, $\Lambda$ has been treated as a weak function of temperature at the walls which are constant values.
Finally, the various properties of the fluid mixture can be expressed as a linear function of \(\phi\) as follows:\\

\begin{itemize}
    \item Density : $\frac{\rho_0}{\rho_2} = \frac{1}{2} \left(1+\rho\right) + \frac{\phi}{2} \left( \rho-1 \right)$ 
    \item Dynamic viscosity : $\frac{\eta^*}{\eta_2} = \frac{1}{2}\left( 1+\eta \right) + \frac{\phi}{2} \left( \eta-1 \right)$ 
    \item Thermal conductivity : $\frac{\kappa^*}{\kappa_2} = \frac{1}{2} \left( 1+\kappa \right) + \frac{\phi}{2} \left( \kappa-1 \right)$ 
    \item Heat capacity : $\frac{\rho_0 c^*}{\rho_2 c_2} = \frac{1}{2}\left(1+ \rho c \right) + \frac{\phi}{2}\left( \rho c - 1 \right)$ \\
\end{itemize} 

\noindent Here, $\rho\;(={\rho_1}/{\rho_2})$ is the density ratio, $\eta\;(={\eta_1}/{\eta_2})$ is the dynamic viscosity ratio, $\beta\;(={\beta_1}/{\beta_2})$ is the ratio of thermal expansion coefficients, $\kappa\;(={\kappa_1}/{\kappa_2})$ is the thermal conductivity ratio, and $c\;(={c_1}/{c_2})$ is specific heat ratio, respectively. While these ratios correspond to the properties of the pure fluids, we also define a few apparent property ratios like $\Tilde{\rho} = {\Tilde{\rho}_1}/{\Tilde{\rho}_2}$, $\Tilde{\beta} = {\Tilde{\beta}_1}/{\Tilde{\beta}_2}$, and $\Tilde{\alpha} = {\Tilde{\alpha}_1}/{\Tilde{\alpha}_2}$, that correspond to the properties of the fluid layers that vary with the miscibility parameter. In fact, each of these ratios approaches unity as the system gets closer to the critical point.

\section{Linear stability analysis}

\begin{figure}
    \centering
    \includegraphics[scale=0.52]{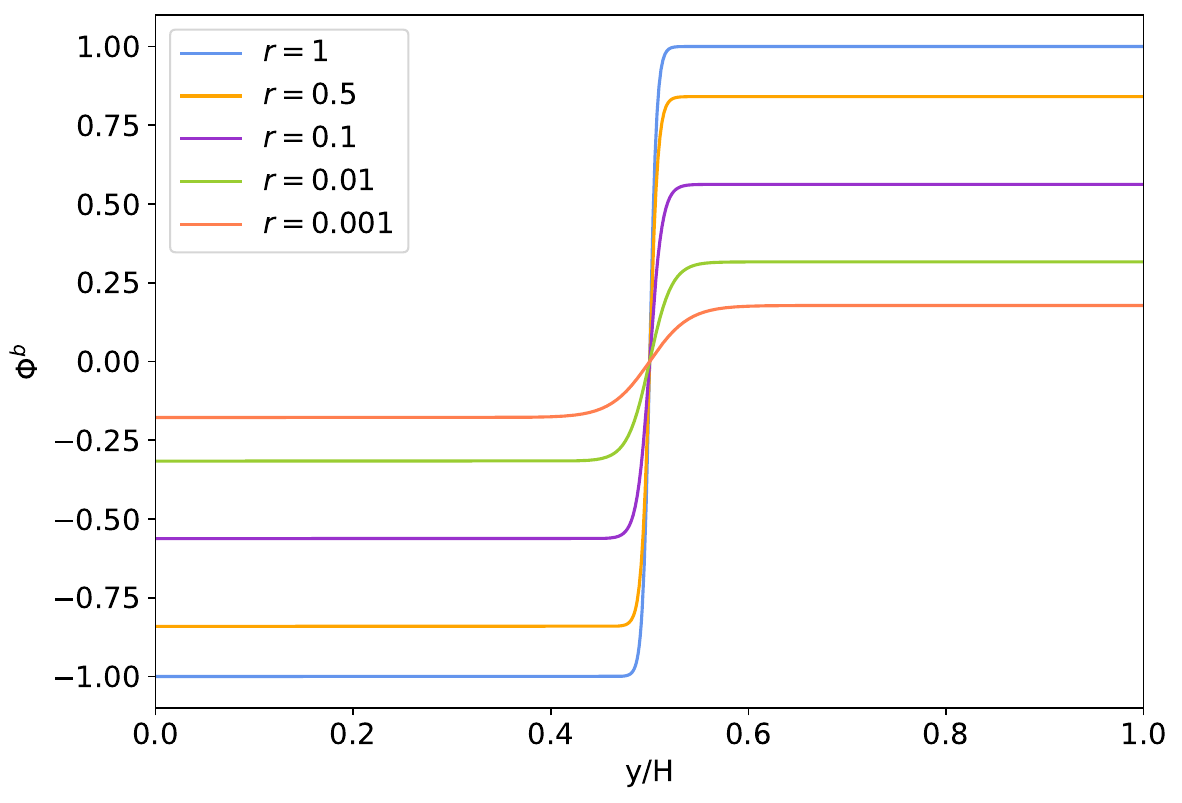}
    \caption{Phase distribution for different $r$}
    \label{fig:phase_distribution_with_r}
\end{figure}

Using the above diffuse interface formulation, we now analyze the onset characteristics of RBM convection in binary fluids closer to their UCST. To this effect, we perform a linear stability analysis to evaluate the critical parameters and the associated modes of convection, which could be either oscillatory or non-oscillatory. Note that the oscillatory modes of onset are observed in a purely immiscible RB scenario when the mechanical and thermal coupling modes compete with each other. This situation of overstability typically occurs when the value of \(\rho\beta\alpha\) differs significantly from unity \citep{Renardy1996}. The present goal is to understand how such oscillatory modes transform for a diffuse interface, particularly when its thickness diverges near the critical point.

The present base state consists of two quiescent layers of fluids with imposed uniform temperatures at the top and bottom walls. Note that the bottom wall's temperature is higher than that of the top and, at the same time, does not exceed the UCST of the mixture. As mentioned earlier, the miscibility transition parameter, `$r$',  is assumed to be uninfluenced by the imposed spatial inhomogeneity in temperature. We accordingly consider a homogeneous $r$ in the domain that corresponds to the interfacial temperature. In fact, $r$ is used here to define the state of the system, and we consider different $r$ values ranging from 0.005 to one, representing the system's closeness (or otherwise) to UCST. Thus, for a given value of $r$, the base profile of the order parameter ($\Phi^b$) can be obtained by considering trivial chemical potential, {\it i.e.} $\mu = 0$. The resulting expression for $\Phi^b$ can be written as  
\begin{equation}
    \Phi ^b(y) = \pm r^\frac{1-q}{2} \tanh \left(\frac{y}{\sqrt{2}\epsilon} r^\frac{1-p}{2} \right) \label{eq:base_Phi}
\end{equation}

Figure~\ref{fig:phase_distribution_with_r} shows the variation of $\Phi^b$ with decreasing `$r$', \textit{i.e.}, with increasing operating temperature. It is evident that the peak value of `$\Phi$' reduces in both the layers owing to the enhanced solubility of the phases. The interfacial width, being inversely proportional to $r^{\frac{p-1}{2}}$, increases with the increase in system temperature. Here, the contribution of the gradient energy term decreases, and more material is introduced into the interfacial region, thereby creating a wider interface.

Along with Eq.~\eqref{eq:base_Phi}, the quiescent base state is represented here through the following expressions for pressure and temperature distribution.
\begin{eqnarray}
     \bnabla p^b(y) + r^p \Lambda(\Theta^b) \nabla^2\Phi^b \bnabla\Phi^b 
     - \frac{1}{3}  r^p \left|\bnabla\Phi^b\right|^2 \frac{\partial\Lambda}{\partial\theta} {\bnabla}\Theta^b \nonumber \\
     + \frac{1}{6} \Lambda(\Theta^b) r^p \bnabla \left|\bnabla\Phi^b\right|^2 + r^p {\bnabla}\Phi^b \frac{\partial\Lambda}{\partial\theta} {\bnabla}\Theta^b\cdot{\bnabla}\Phi^b& = & - \rho ^{'} \left(\Phi^b, \Theta^b\right)g \boldsymbol{1}_y \\
      \bnabla \cdot \left(\kappa (\Phi^b) \bnabla \Theta^b(y)\right) & = & 0
\end{eqnarray}
\begin{align*}
    L_R& = H  &  t_R& = \frac{{H}^2}{\alpha_2}   &   u_R& = \frac{\alpha_2}{H}   &    \theta_R& = \theta_B - \theta_T     &   p_R& = \frac{\rho_2 \alpha_2 \nu _2}{{H}^2}
\end{align*}
\noindent The non-dimensionalization of all the variables has been performed in the current formulation using the above scales that correspond to the top layer. $H$ is the total height of the system. Correspondingly, we arrive at four relevant dimensionless numbers, which are 1) the non-dimensional mobility, $M\;(= \gamma\Lambda_0/\epsilon^2\alpha_2)$, which is the ratio of the interfacial diffusivity to the thermal diffusivity of the reference fluid, 2) the modified inverse Capillary number, $\Gamma_\theta\;(= \Lambda_0/\rho _2 \alpha_2^2)$, where the word 'modified' refers to the non-standard usage of thermal diffusivity instead of the kinematic viscosity, 3) the inverse capillary number, $Ca \;(= \sigma H/\eta\alpha_2)$, and 4) the Marangoni number, $Ma\;(=\sigma_\theta \Delta\theta H/\eta_2\alpha_2)$. Since the Marangoni $(Ma)$ and the Rayleigh numbers, $Ra\;(= g \rho_2 \beta_2 \Delta\theta H^3/\eta_2\alpha_2)$, are not independent, an additional non-dimensional parameter, $\zeta$, has been defined (below) to link the two.
\begin{equation}
\zeta = \frac{Ma}{Ra} = \frac{\sigma_\theta}{g\rho_2\beta_2 H^2}
\end{equation}
Linearizing the governing equations (Eq.~\eqref{feq_phi} to Eq.~\eqref{feq_ener}) and the boundary conditions over the above base state, we obtain the ensuing set of equations that govern the evolution of perturbations in the domain.
\begin{equation}
    \frac{\partial \phi^{'}}{\partial t} + {u}_j^{'}{\nabla}_j\Phi^b 
    = M \nabla^2 \left[ \left(1 - \frac{\zeta Ra}{Ca} \Theta^{b}\right) \left( r^q 3(\Phi^b)^2\phi^{'} - r \phi^{'} - \left( \frac{\epsilon}{H} \right)^2 r^p \nabla^2\phi^{'} \right) \right]
\end{equation}

\begin{eqnarray}
    {\rho_f^b}\frac{\partial {u}_i^{'}}{\partial t} 
    &= &- Pr_2 {\nabla}_i p^{'} + Pr_2 {\nabla}_j \left[ \left( \frac{1}{2}(1+\eta) + \frac{\Phi^b}{2}(\eta-1) \right) \left( {\nabla}_j u_i^{'} + {\nabla}_i u_j^{'} \right) \right]  \nonumber \\
    &&+ Ra Pr_2 \frac{1}{2}\left(\rho\beta-1\right) \Theta^b \phi^{'} \delta_{i2} + Ra Pr_2 \left[\frac{1}{2}\left(\rho\beta+1\right) + \frac{1}{2}\left(\rho\beta-1\right) \Phi^{b} \right] \theta^{'} \delta_{i2}  \nonumber \\
    &&+ \left(\frac{H}{\epsilon}\right)^2 \Gamma_\theta \left( r^q 3(\Phi^b)^2 \phi^{'} - r \phi^{'} - \left(\frac{\epsilon}{H}\right) ^2 r^p \nabla^2 \phi^{'} \right)\nabla_i\Phi^{b} \nonumber \\
    &&- \frac{3}{2\sqrt{2}} \zeta Ra Pr_2 \frac{H}{\epsilon} \left[\Theta^b \left( r^q 3(\Phi^b)^2 \phi^{'} - r \phi^{'} - \left(\frac{\epsilon}{H}\right) ^2 r^p \nabla^2 \phi^{'} \right)\nabla_i\Phi^{b} \right] \nonumber \\
    &&- \frac{3}{2\sqrt{2}} \zeta Ra Pr_2 \frac{\epsilon}{H} r^p \nabla_i\Theta^b \nabla^2\Phi^b \phi^{'} \nonumber \\
    &&+ \frac{3}{2\sqrt{2}} \zeta Ra Pr_2 \frac{\epsilon}{H} r^p \frac{1}{3} \left[\Theta^b \left(\partial_y\Phi^b \nabla_i(\partial_y\phi^{'}) + \nabla_i(\partial_y\Phi^b) \partial_y\phi^{'}\right) \right] \nonumber \\
    &&+ \frac{3}{2\sqrt{2}} \zeta Ra Pr_2 \frac{\epsilon}{H} r^p \frac{1}{3} \partial_y\Phi^b \nabla_i\left(\partial_y\Phi^b\right) \theta^{'}  \nonumber \\
    &&+ \frac{3}{2\sqrt{2}} \zeta Ra Pr_2 \frac{\epsilon}{H} r^p \nabla^2\Phi^b\nabla_i\Phi^b\theta^{'} \nonumber \\
    &&- \frac{3}{2\sqrt{2}} \zeta Ra Pr_2 \frac{\epsilon}{H} r^p \left( \frac{1}{3} \left|\nabla_k\Phi^b\right|^2 \nabla_i\theta^{'} + \frac{2}{3} \nabla_k\Phi^b \nabla_k\phi^{'} \nabla_i\Theta^b\right)    \\
    &&+ \frac{3}{2\sqrt{2}} \zeta Ra Pr_2 \frac{\epsilon}{H} r^p \left(\nabla_i\Phi^b \left(\nabla_j\Theta^b \nabla_j\phi^{'} + \nabla_j\theta^{'} \nabla_j\Phi^b \right) + \nabla_i\phi^{'} \nabla_j\Theta^b \nabla_j\Phi^b \right) \nonumber
\end{eqnarray}

\begin{equation}
    {\nabla}_j {u}_j^{'} = 0
\end{equation}

\begin{eqnarray}
    (\rho c)_f^b\left[\frac{\partial \theta ^{'}}{\partial t} + {u_j}^{'} \frac{\partial \Theta^{b}}{\partial x_j} \right]
    & = & \kappa_f(\Phi^b){\nabla_m}^2\theta ^{'} + (\nabla_m \kappa_f(\Phi^b))(\nabla_m \theta^{'}) \nonumber \\
    &&+ (\nabla_m \kappa_f(\phi^{'})) (\nabla_m \Theta^b) + \kappa_f(\phi^{'}){\nabla_m}^2\Theta^{b} 
\end{eqnarray}

\noindent In the above equations, the property ratios, $(\rho c)_f^b$ and $\rho_f^b$, are defined as 

\begin{eqnarray}
(\rho c)_f^b & = & \frac{1}{2}(1+\frac{\rho_1 c _1}{\rho_2 c _2}) + \frac{\Phi^b}{2}(\frac{\rho_1 c _1}{\rho_2 c _2}-1)    \\
\rho_f^b & = & \frac{1}{2}(1+\rho) + \frac{\Phi^b}{2}(\rho-1) 
\end{eqnarray}

For the current system configuration of infinite horizontal extents, the normal mode expansion of the perturbed quantities can be written as
\begin{equation}\label{eq:normal_mode_expansion_form}
\left\{\phi^{'}, \boldsymbol{u}^{'}, \theta^{'}, p^{'}\right\}\left(x,y,t\right) = \left\{\Hat{\phi}(y), \Hat{\boldsymbol{u}}(y), \Hat{\theta}(y), \Hat{p}(y)\right\}\exp{(\lambda t + \mathrm{i}kx)},
\end{equation}
where $k^2 = k_1^2+k_2^2$ and $k_1$ \& $k_2$ are the wavenumbers of the imposed disturbances along the two orthogonal horizontal directions. Here, $\Hat{\boldsymbol{u}} = \left(\Hat{u}, \Hat{w}\right)$, and $\Hat{u} = \frac{k_1\Hat{u}_1 + k_2\Hat{u}_2}{k}$, $\Hat{u}_1$ and $\hat{u}_2$ are the amplitudes of the horizontal velocity components. Correspondingly, all the linearized perturbation equations get transformed as follows:

\begin{eqnarray}
   \lambda \hat{\phi} + \hat{v} \frac{\mathrm{d} \Phi^{b}}{\mathrm{d} y}
    & = & M \left[  3 r^q \left( -(\Phi^b)^2 k^2 \hat{\phi} + 2 (\frac{\mathrm{d}\Phi^{b}}{\mathrm{d}y})^2 \hat{\phi} + 2 \Phi^b \frac{\mathrm{d}^2\Phi^b}{\mathrm{d}y^2} \hat{\phi} + 4\Phi^b \frac{\mathrm{d}\Phi^b}{\mathrm{d}y} \frac{\mathrm{d}\hat{\phi}}{\mathrm{d}y} + (\Phi^b)^2 \frac{\mathrm{d}^2\hat{\phi}}{\mathrm{d}y^2} \right) \right. \nonumber \\ 
    &&\left. - r \left( -k^2 \hat{\phi} + \frac{\mathrm{d}^2\hat{\phi}}{\mathrm{d}y^2} \right) - \left(\frac{\epsilon}{H}\right) ^2 r^p \left( \frac{\mathrm{d}^4\hat{\phi}}{\mathrm{d}y^4} - 2 k^2 \frac{\mathrm{d}^2\hat{\phi}}{\mathrm{d}y^2} + k^4 \hat{\phi}\right)  \right] \nonumber \\
    &&- M \left( \frac{\zeta Ra}{Ca}  \right) \Theta^b \left[ 3 r^q ( -(\Phi^b)^2 k^2 \hat{\phi} + 2 \left(\frac{\mathrm{d}\Phi^{b}}{\mathrm{d}y}\right)^2 \hat{\phi} + 2 \Phi^b \frac{\mathrm{d}^2\Phi^b}{\mathrm{d}y^2} \hat{\phi} + 4\Phi^b \frac{\mathrm{d}\Phi^b}{\mathrm{d}y} \frac{\mathrm{d}\hat{\phi}}{\mathrm{d}y} \right. \nonumber  \\
    &&\left. + (\Phi^b)^2 \frac{\mathrm{d}^2\hat{\phi}}{\mathrm{d}y^2} ) - r \left( -k^2 \hat{\phi} + \frac{\mathrm{d}^2\hat{\phi}}{\mathrm{d}y^2} \right) - \left(\frac{\epsilon}{H}\right) ^2 r^p \left( \frac{\mathrm{d}^4\hat{\phi}}{\mathrm{d}y^4} - 2 k^2 \frac{\mathrm{d}^2\hat{\phi}}{\mathrm{d}y^2} + k^4 \hat{\phi}\right) \right]  \nonumber  \\
    &&- 2 M \left( \frac{\zeta Ra}{Ca}  \right)  \frac{\mathrm{d}\Theta^b}{\mathrm{d}y} \left[ 3 r^q(\Phi^b)^2 \frac{\mathrm{d}\hat{\phi}}{\mathrm{d}y} + 6 r^q(\Phi^b) \frac{\mathrm{d}\Phi^b}{\mathrm{d}y} \hat{\phi} - r \frac{\mathrm{d}\hat{\phi}}{\mathrm{d}y} - \left( \frac{\epsilon}{H} \right)^2 r^p \left( \frac{\mathrm{d}^2}{\mathrm{d}y^2} - k^2 \right) \frac{\mathrm{d}\hat{\phi}}{\mathrm{d}y} \right] \nonumber \\
    &&- M \left( \frac{\zeta Ra}{Ca}  \right) \frac{\mathrm{d}^2\Theta^b}{\mathrm{d}y^2} \left( r^q 3(\Phi^b)^2 \Hat{\phi} - r \Hat{\phi} - \left( \frac{\epsilon}{H} \right)^2 r^p \nabla^2\Hat{\phi} \right) \label{eq:disc_ch}
\end{eqnarray}

\begin{eqnarray}
   {\rho_f(\Phi^b)} \lambda\hat{u} &= &- Pr_2 (\mathrm{i} k) \hat{p} 
    + Pr_2 \left[ \frac{1}{2}(1+\eta) + \frac{\Phi^b}{2}(\eta-1) \right] \left(-k^2+\frac{\mathrm{d}^2}{\mathrm{d}y^2}\right)\hat{u} \nonumber \\
    && + Pr_2 \frac{1}{2}(\eta-1) \frac{\mathrm{d}\Phi^b}{\mathrm{d}y}\frac{\mathrm{d}\hat{u}}{\mathrm{d}y} + Pr_2 \frac{1}{2}(\eta-1) \frac{\mathrm{d}\Phi^b}{\mathrm{d}y}(\mathrm{i}k \hat{v}) \nonumber \\
    && -  \frac{3}{2\sqrt{2}} \zeta Ra Pr_2  \frac{\epsilon}{H} \frac{1}{3} r^p \left( \frac{\mathrm{d}\Phi^b}{\mathrm{d} y} \right) ^2 \left( \mathrm{i}k\Hat{\theta} \right) + \frac{3}{2\sqrt{2}} \zeta Ra Pr_2 \frac{\epsilon}{H} r^p \frac{\mathrm{d}\Phi^{b}}{\mathrm{d} y} \frac{\mathrm{d}\Theta^b}{\mathrm{d} y} \left( \mathrm{i}k\Hat{\phi} \right) \nonumber \\ \label{eq:disc_u}
\end{eqnarray}

\begin{eqnarray}
    {\rho_f(\Phi^b)} \lambda\hat{v} &= &-  Pr_2 \frac{\mathrm{d}\hat{p}}{\mathrm{d}y} 
    + Pr_2 \left[ \frac{1}{2}(1+\eta) + \frac{\Phi^b}{2}(\eta-1) \right] \left(-k^2+\frac{\mathrm{d}^2}{\mathrm{d}y^2}\right)\hat{v} \nonumber \\
    && + Pr_2 (\eta-1) \frac{\mathrm{d}\Phi^b}{\mathrm{d}y}\frac{\mathrm{d}\hat{v}}{\mathrm{d}y} + Ra Pr_2 \frac{1}{2}\left(\rho\beta-1\right) \Theta^b \hat{\phi} \nonumber \\
    && +  Ra Pr_2           \left[\frac{1}{2}\left(\rho\beta+1\right) + \frac{1}{2}\left(\rho\beta-1\right) \Phi^{b} \right] \hat{\theta} \nonumber \\
    &&
    + \left(\frac{H}{\epsilon}\right)^2 \Gamma_\theta \left( r^q 3(\Phi^b)^2 \Hat{\phi} - r \Hat{\phi} - \left(\frac{\epsilon}{H}\right) ^2 r^p \left(-k^2+\frac{\mathrm{d}^2}{\mathrm{d}y^2}\right) \hat{\phi} \right)\frac{d\Phi^b}{dy} \nonumber \\ 
    &&- \frac{3}{2\sqrt{2}} \zeta Pr_2 Ra \frac{H}{\epsilon} \Theta^{b} \left( r^q 3(\Phi^b)^2 \Hat{\phi} - r \Hat{\phi} - \left(\frac{\epsilon}{H}\right) ^2 r^p \left(-k^2+\frac{\mathrm{d}^2}{\mathrm{d}y^2}\right) \hat{\phi} \right)\frac{d\Phi^b}{dy}  \nonumber \\
    &&
    -  \frac{3}{2\sqrt{2}} \zeta Ra Pr_2 \frac{\epsilon}{H} \frac{\mathrm{d}\Theta^b}{\mathrm{d}y} r^p \frac{\mathrm{d}^2\Phi}{\mathrm{d}y^2} \hat{\phi} 
    +  \frac{3}{2\sqrt{2}} \zeta Ra Pr_2 \frac{\epsilon}{H} \frac{4}{3} r^p \left(\frac{\mathrm{d}^2\Phi^{b}}{\mathrm{d} y^2} \frac{\mathrm{d}\Phi^b}{\mathrm{d} y} \right) \hat{\theta} \nonumber \\
    &&
    +  \frac{3}{2\sqrt{2}} \zeta Pr_2 Ra \frac{\epsilon}{H} \frac{2}{3} r^p \left( \frac{\mathrm{d}\Phi^b}{\mathrm{d} y} \right) ^2  \frac{\mathrm{d}\Hat{\theta}}{\mathrm{d}y} 
    + \frac{3}{2\sqrt{2}} \zeta Ra Pr_2 \frac{\epsilon}{H} r^p \frac{\mathrm{d}\Phi^{b}}{\mathrm{d} y} \frac{\mathrm{d}\Theta^b}{\mathrm{d} y} \frac{\mathrm{d}\Hat{\phi}}{\mathrm{d}y} \nonumber \\ \label{eq:disc_v}
\end{eqnarray}

\begin{equation}
    \mathrm{i}k \hat{u} + \frac{\mathrm{d} \hat{v}}{\mathrm{d} y} = 0  \label{eq:disc_conti}
\end{equation}

\begin{eqnarray}
   (\rho c)_f (\Phi^b)\left[ \lambda \hat{\theta} + \hat{v} \frac{\mathrm{d} \Theta^{b}}{\mathrm{d} y} \right] &
    = & \left[ \frac{1}{2}(1+\kappa) + \frac{\Phi^b}{2}(\kappa-1) \right]\left(-k^2+\frac{\mathrm{d}^2}{\mathrm{d} y^2}\right)\hat{\theta}  \nonumber \\
    && +\frac{1}{2}(\kappa-1) \frac{\mathrm{d} \Phi^b}{\mathrm{d} y}\frac{\mathrm{d} \hat{\theta}}{\mathrm{d} y} + \frac{1}{2}(\kappa-1)\frac{\mathrm{d} \hat{\phi}}{\mathrm{d} y}\frac{\mathrm{d} {\Theta^b}}{\mathrm{d} y}  \nonumber \\
    && + \frac{\Hat{\phi}}{2}(\kappa-1) \frac{\mathrm{d}^2 \Theta^b}{\mathrm{d} y^2} \label{eq:disc_temp}
\end{eqnarray}

The boundary conditions at the top and bottom plates are written as
\begin{equation}
    \begin{split}
     \frac{\mathrm{d}\hat{\phi}}{\mathrm{d} y} &=  \frac{\mathrm{d}^3\hat{\phi}}{\mathrm{d} y^3} = 0 \,\,\,\,\,\,@ y = 0, 1  \\ 
     \hat{u}_i &= 0 \,\,\,\,\,\,\,\,\,@ y = 0, 1 \\
     \hat{\theta} &= 0 \,\,\,\,\,\,\,\,\,@ y = 0, 1 \\
    \end{split} \label{eq:bouncon}
\end{equation}

The dependent variables in the above set of semi-discrete equations are further expanded using the Chebyshev collocation method along the vertical coordinate. Here, any variable \(\psi\) is expressed via a Lagrangian interpolation of values at the collocation points as
\begin{equation}
    \psi_N (y) = \sum_{i=0}^{N} h_i (y) \psi (y_i),
\end{equation}
where \(N\) is the number of collocation points along the vertical direction. The cardinal function, \(h_i (y)\), defined over the collocation points, \(y_i=\cos\left({\frac{\pi N}{i}}\right)\), is expressed as 
\begin{equation}
    h_i (y) = \frac{(-1)^{i+1}(1-y^2){T_N}^{'}(y)}{{\overline{c}_i}N^2(y-y_i)},
\end{equation}
where \(\overline{c}_i=2\) for the endpoints and \(1\) for all the interior points. \({T_N}^{'}(y)\) is the $N^{th}$ derivative of the Chebyshev polynomial of the first kind. To avoid the generation of spurious pressure modes, we use the $P_N - P_{N-2}$ formulation wherein the pressure variable alone is expanded as 
\begin{equation}
    p_{N-2} (y) = \sum_{i=1}^{N-1} \hat{h}_i (y) p(y_i).
\end{equation}
The cardinal function, $\hat{h}_i (y)$, for the above pressure expansion is defined as
\begin{equation}
    \hat{h}_i (y) = \frac{(1-{y_i}^2)}{(1-y)} h_i (y)
\end{equation}

In the standard spectral collocation method, the choice of grid points is often restricted to the Gauss-Lobatto-Chebyshev (G-L-C) points owing to the stringent accuracy requirements. These G-L-C points are finer at the domain extremities and are coarser in the middle. Unfortunately, such a configuration poses an issue in the present scenario, where sufficient grid points are needed around the diffuse interface to calculate the gradients of all the variables accurately. In order to mitigate this issue, one might consider using an exorbitant number of points so that the diffuse interface is sufficiently resolved. Unfortunately, such mesh refinement often makes the solution process time-consuming. In the present work, we use an alternate means of resolving the diffuse interface through grid mapping. The grid points clustered around the diffuse interface in the physical domain are suitably mapped to the classical G-L-C points in the computational domain. In this regard, we use a transformation suggested by \citet{Tee2006} as given below.
\begin{equation}
    \Tilde{y} = g(y) = \delta + \in \sinh \left[ \left({ \arcsinh \left(\frac{1-\delta}{\in}\right) } + \arcsinh \left(\frac{1+\delta}{\in}\right) \right) \frac{y-1}{2} + \arcsinh \left(\frac{1-\delta}{\in}\right) \right]
\end{equation} 
Here, \(\in\) decides the arrangement of the nodes in the transformed grid system, and the value of \(\delta\) determines the vertical position of the interface in the domain. Lower the value of \(\in\), denser the interfacial region in the mapped domain (refer to Fig.2 of \cite{Diwakar2015}). Note that \(\in\) has to be chosen carefully as a dense interfacial region would starve other regions of points and may result in significant errors while calculating the higher-order derivatives. The relations used presently to transform the spatial derivatives from the G-L-C grid to the physical domain have been provided in Appendix \ref{appA}.

With the above polynomial expansions and grid transformations, the discrete version of perturbation equations can be written as a Generalized Eigenvalue Problem (GEP) of the form, \(\boldsymbol{AX}=\lambda \boldsymbol{BX}\). Here, one can deploy two means for identifying the critical $Ra$ for onset. In the first approach, the GEP can be formulated in such a way that $Ra$ directly becomes the eigenvalue being evaluated. Alternatively, an iterative procedure can be used to search for the lowest $Ra$ at which the real part of the largest eigenmode is zero, i.e., $Real(\lambda)=0$. In any case, solving the GEP directly in the above form may give rise to spurious modes due to the zero-valued rows in matrix $\boldsymbol{B}$ corresponding to the continuity equation and the boundary conditions. These spurious modes pose a serious threat as they could often be confused with the desired eigenvalues. To avoid such a scenario, we employ the reciprocal approach wherein we seek the eigenvalues of the system, \(\Tilde{\lambda} \boldsymbol{AX} = \boldsymbol{BX}\). Subsequently, filtration of the obtained eigenvalues is carried out by ignoring those corresponding to the trivial eigenvectors. In the present work, the full spectra of eigenvalues of the reciprocal system are evaluated using the standard QR method (EIG) of the GNU-Octave package. Upon sorting and inverting these filtered eigenvalues, we arrive at the solution for the original GEP, which helps provide information on both the critical parameters $(Ra)$ for flow onset and the nature of convection. Here, a non-trivial imaginary part of the leading eigenvalue helps identify an oscillatory onset of convection in the system.
 
\section{RB convection in binary fluids}
In the present analysis, we first consider the scenario of RB convection, wherein the instability is driven by buoyancy effects in one or both layers. Here, the variation of interfacial tension with temperature or any other parameter is neglected ($\zeta = 0$). Correspondingly, the semi-discrete equations governing the evolution of $\phi$, $u$-momentum, and $v$-momentum, i.e. Eqs.~\eqref{eq:disc_ch}, \eqref{eq:disc_u}, and \eqref{eq:disc_v}, gets modified as

\begin{eqnarray}
   \lambda \hat{\phi} + \hat{v} \frac{\mathrm{d} \Phi^{b}}{\mathrm{d} y}
    & = & M \left[  3 r^q \left( -(\Phi^b)^2 k^2 \hat{\phi} + 2 \left(\frac{\mathrm{d}\Phi^{b}}{\mathrm{d}y}\right)^2 \hat{\phi} + 2 \Phi^b \frac{\mathrm{d}^2\Phi^b}{\mathrm{d}y^2} \hat{\phi} + 4\Phi^b \frac{\mathrm{d}\Phi^b}{\mathrm{d}y} \frac{\mathrm{d}\hat{\phi}}{\mathrm{d}y} + (\Phi^b)^2 \frac{\mathrm{d}^2\hat{\phi}}{\mathrm{d}y^2} \right) \right. \nonumber \\ 
    &&\left. - r \left( -k^2 \hat{\phi} + \frac{\mathrm{d}^2\hat{\phi}}{\mathrm{d}y^2} \right) - \left(\frac{\epsilon}{H}\right) ^2 r^p \left( \frac{\mathrm{d}^4\hat{\phi}}{\mathrm{d}y^4} - 2 k^2 \frac{\mathrm{d}^2\hat{\phi}}{\mathrm{d}y^2} + k^4 \hat{\phi}\right)  \right] \label{eq:disc_ch1}
\end{eqnarray}

\begin{eqnarray}
    \lambda\hat{u} &= &- \frac{1}{\rho_f(\Phi^b)} Pr_2 (\mathrm{i} k) \hat{p} 
    + \frac{1}{\rho_f(\Phi^b)} Pr_2 \left[ \frac{1}{2}(1+\eta) + \frac{\Phi^b}{2}(\eta-1) \right] \left(-k^2+\frac{\mathrm{d}^2}{\mathrm{d}y^2}\right)\hat{u} \nonumber \\
    && + \frac{1}{\rho_f(\Phi^b)} Pr_2 \frac{1}{2}(\eta-1) \frac{\mathrm{d}\Phi^b}{\mathrm{d}y}\frac{\mathrm{d}\hat{u}}{\mathrm{d}y} + \frac{1}{\rho_f(\Phi^b)} Pr_2 \frac{1}{2}(\eta-1) \frac{\mathrm{d}\Phi^b}{\mathrm{d}y}(\mathrm{i}k \hat{v})
    \label{eq:disc_u1}
\end{eqnarray}

\begin{eqnarray}
    \lambda\hat{v} &= &- \frac{1}{\rho_f(\Phi^b)} Pr_2 \frac{\mathrm{d}\hat{p}}{\mathrm{d}y} 
    + \frac{1}{\rho_f(\Phi^b)} Pr_2 \left[ \frac{1}{2}(1+\eta) + \frac{\Phi^b}{2}(\eta-1) \right] \left(-k^2+\frac{\mathrm{d}^2}{\mathrm{d}y^2}\right)\hat{v} \nonumber \\
    && + \frac{1}{\rho_f(\Phi^b)} Pr_2 (\eta-1) \frac{\mathrm{d}\Phi^b}{\mathrm{d}y}\frac{\mathrm{d}\hat{v}}{\mathrm{d}y} 
    + \frac{1}{\rho_f(\Phi^b)} Ra Pr_2 \frac{1}{2}\left(\rho\beta-1\right) \Theta^b \hat{\phi} \nonumber \\
    && + \frac{1}{\rho_f(\Phi^b)} Ra Pr_2           \left[\frac{1}{2}\left(\rho\beta+1\right) + \frac{1}{2}\left(\rho\beta-1\right) \Phi^{b} \right] \hat{\theta} \nonumber \\
    &&
    + \left(\frac{H}{\epsilon}\right)^2 \Gamma_\theta \frac{1}{\rho_f(\Phi^b)} \left( r^q 3(\Phi^b)^2 \Hat{\phi} - r \Hat{\phi} - \left(\frac{\epsilon}{H}\right) ^2 r^p \left(-k^2+\frac{\mathrm{d}^2}{\mathrm{d}y^2}\right) \hat{\phi} \right)\frac{d\Phi^b}{dy} 
    \label{eq:disc_v1}
\end{eqnarray}

Along with the above three simplified equations, the semi-discrete continuity equation (Eq.\eqref{eq:disc_conti}) and the equation governing the evolution of thermal perturbations (Eq.\eqref{eq:disc_temp}) are simultaneously solved to characterise the onset of RB convection in the system. The boundary conditions remain the same as in Eq.\eqref{eq:bouncon}. Owing to the present `one-fluid' formulation, no explicit interfacial conditions are specified here and hence, neglecting the Marangoni effect does not modify any of the associated conditions.

\subsection{Choice of fluids and operating conditions}
Thermo-convective flows in multi-layer systems are characterised by numerous non-dimensional parameters, even in their simplest setting \citep{Diwakar2014JFM}. Hence, identifying the parametric space for oscillatory convection in such systems becomes challenging. This task becomes even more strenuous in the present phase-field formulation as four additional non-dimensional parameters are utilised to characterise the convection. As a remedy, we resort to the concept of balanced contrast proposed by \citet{Colinet1994} wherein the fluids are chosen such that the combination of properties, $\rho\beta\alpha$, is maintained at a constant value. At the same time, following \citet{Diwakar2014JFM}, we choose the other properties such that they yield unique $a^{*}$ values. $a^{*}$ is the critical height ratio at which the Rayleigh numbers of the two layers are equal and is given as 
\begin{equation}
    a^{*} = \left(\frac{\beta \rho}{\kappa{\alpha}\eta}\right)^{\frac{1}{4}}.
\end{equation}

The following property ratios have been chosen here to fix the value of $\rho\beta\alpha$ combination to be 0.125.
\begin{align*}
    \rho = \frac{\rho_1}{\rho_2} & = 2 & \eta = \frac{\eta_1}{\eta_2} & = 1 &
    \beta = \frac{\beta_1}{\beta_2} & = 0.125 &
    \kappa = \frac{\kappa_1}{\kappa_2} & = 0.5 & 
    c = \frac{c_1}{c_2} & = 0.5 &
\end{align*}

Three different values of $a^{*}$ $(= 1.0, 0.667, 1.5)$ have been considered for the present analysis, and they have been obtained by choosing the viscosity ratio to be $1.0$, $5.0625$, and $0.1975$, respectively. Note that the fluid mixture at the top layer should always be lighter than the bottom layer to avoid the manifestation of Rayleigh-Taylor instability. Hence, the thermal expansivity of the top layer has been chosen to be relatively higher to prevent the overturning of the fluids. 

The other relevant non-dimensional parameters utilized here are $M = 1500$, $\Gamma_\theta = 1000$, and $Pr_2 = 1.0$. As mentioned before, the operating condition of the system, i.e. the extent of its closeness to the UCST, is chosen by independently varying the miscibility transition parameter, $r$, which is a surrogate for the base system's mean temperature. $r$ is varied between 1.0 and 0.005, wherein $r=1.0$ implies that the fluid pair is in an ideal immiscible state and $r=0.005$ corresponds to operating conditions closer to UCST. 

\subsection{Check for consistency}

\begin{figure}
    \centering
    \includegraphics[scale=0.44]{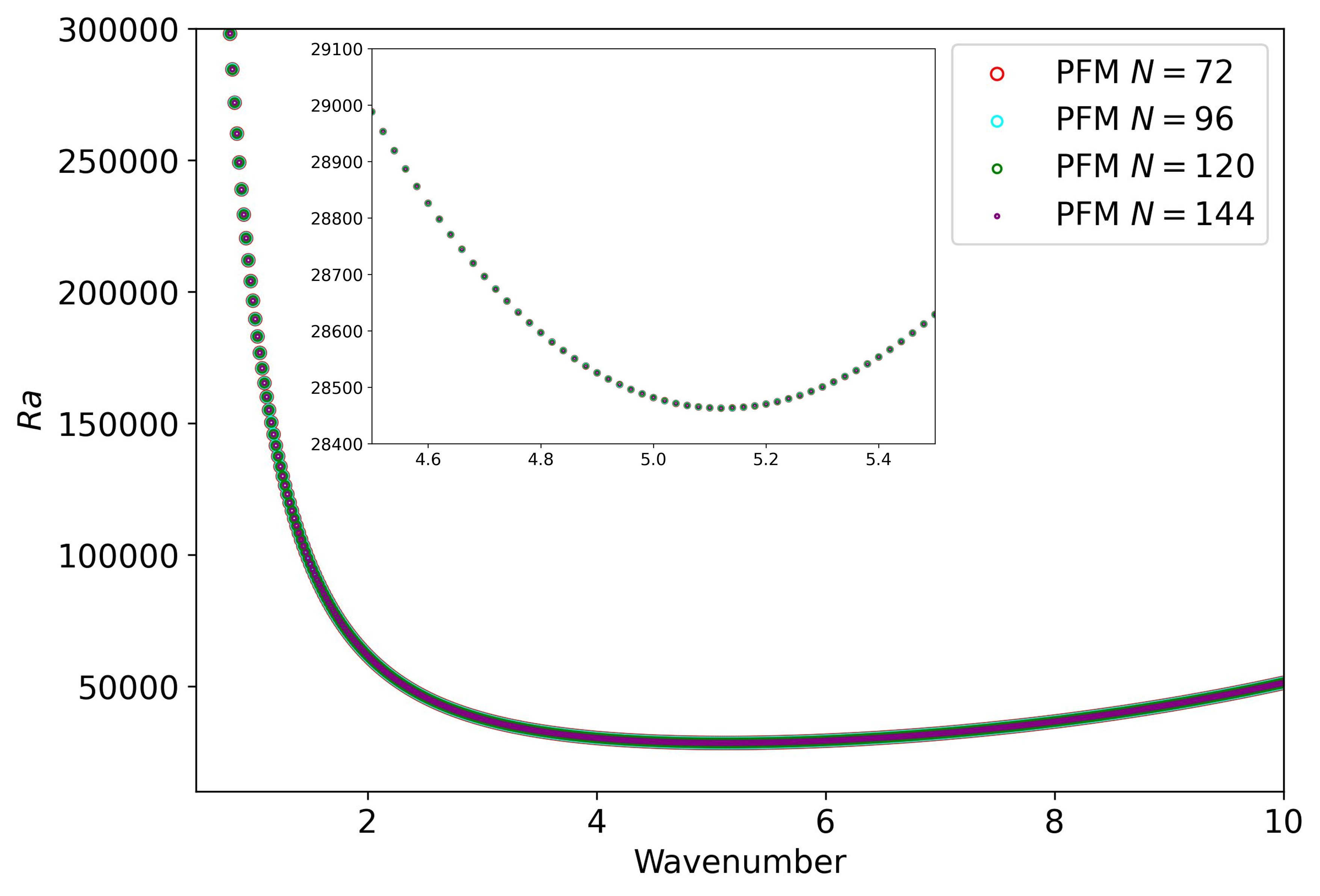}   \caption{Critical Rayleigh numbers ($Ra$) vs the wavenumber ($k$) of the applied disturbances for a fluid system having $\rho\beta\alpha = 0.125$, $a^* = 1.0$, $r=1$, and an interfacial height of \(y_I = 0.5\)}
    \label{fig:grid_independence_graph}
\end{figure}
From the correctness/consistency perspective, the current PFM should ideally reproduce the sharp interface results in the limit of vanishing interfacial thickness. This consistency is now verified by comparing the corresponding results with those of the DDM implementation by \citet{Diwakar2014JFM}. However, as a precursor, one must first estimate the optimum number of G--L--C points for the calculations. Since a mapping strategy has been deployed to accurately estimate the interfacial gradients, more G--L--C points would be typically required to arrive at the mesh-independent results. This is evident from Fig.~\ref{fig:grid_independence_graph} where the critical Ra has been plotted against the wavenumber of perturbations for a fluid system with $\rho\beta\alpha = 0.125$, $a^* = 1.0$, $r=1$, and an interfacial height of \(y_I = 0.5\). The figure shows identical results for grid counts greater than 72 at $\in = 0.0001$, and the DDM produces the same results with a lower grid count of 24. Understandably, an increased computational effort is associated with the present PFM compared to the DDM approach. However, this increased cost pales in comparison to the benefits of the present scheme, i.e., its ability to characterise RBM convection in systems with diffuse interfaces. Since the results obtained for grid counts beyond 72 are mesh-independent, all the calculations henceforth use 72 G--L--C points.  

\begin{figure}
\centering
\includegraphics[scale=0.52]{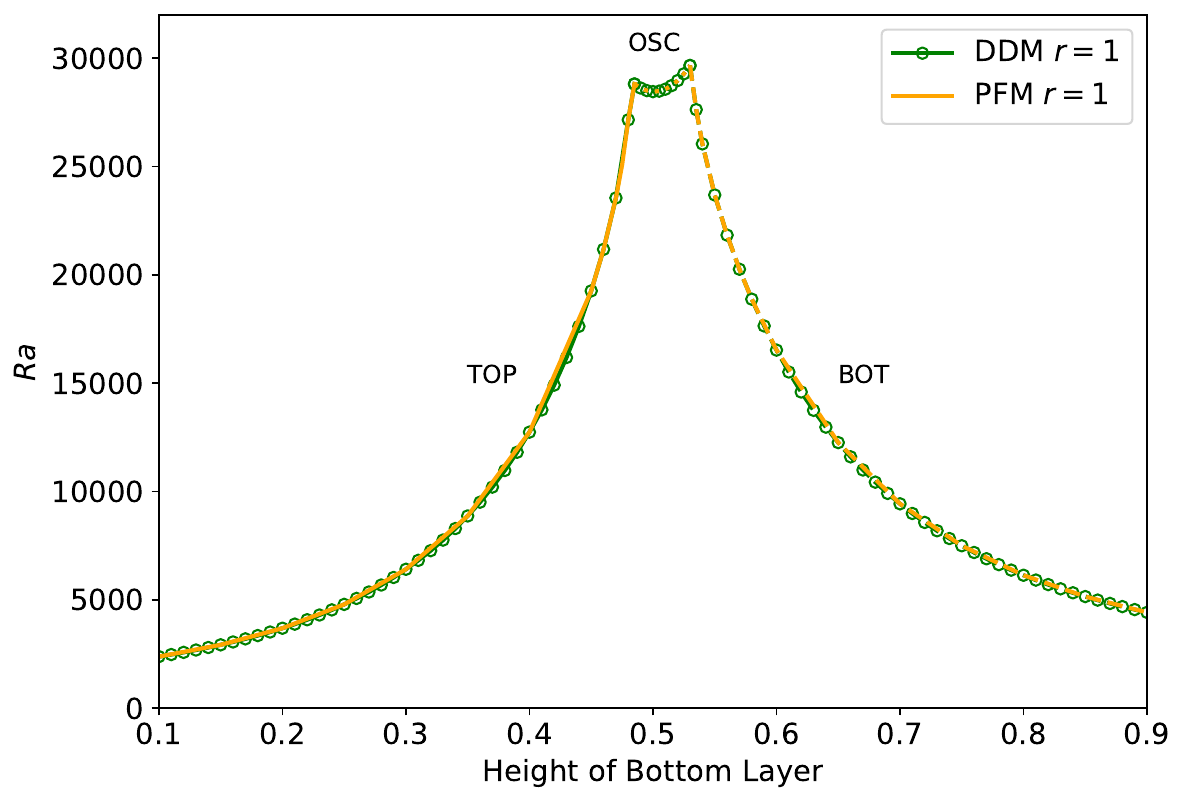} 
\label{fig:RB_neutral_curve_DDM_vs_PFM_complete}
\caption{A comparison between the neutral curves obtained from domain decomposition method (DDM) and phase-field method (PFM) for immiscible fluids}
\label{fig:RB_neutral_curve_DDM_vs_PFM}
\end{figure}

Proceeding further, we perform the consistency check wherein the sharp interface consideration is realised in the present PFM by choosing the value of ${\epsilon}/{H}$ to be $10^{-4}$. Figure~\ref{fig:RB_neutral_curve_DDM_vs_PFM} shows the critical Ra versus the interfacial height plot for the fluid system with $\rho\beta\alpha = 0.125$, $r = 1.0$, and $a^* = 1.0$. The critical Ra at each interfacial height corresponds to the lowest critical Ra values obtained for different input wavenumbers. It is evident from the figure that an exact match is obtained between the DDM and the phase field approaches for the prescribed conditions. The figure shows three notable regimes, whose significance will be described in the ensuing section. 

\begin{figure}
    \centering
    \includegraphics[scale=0.52]{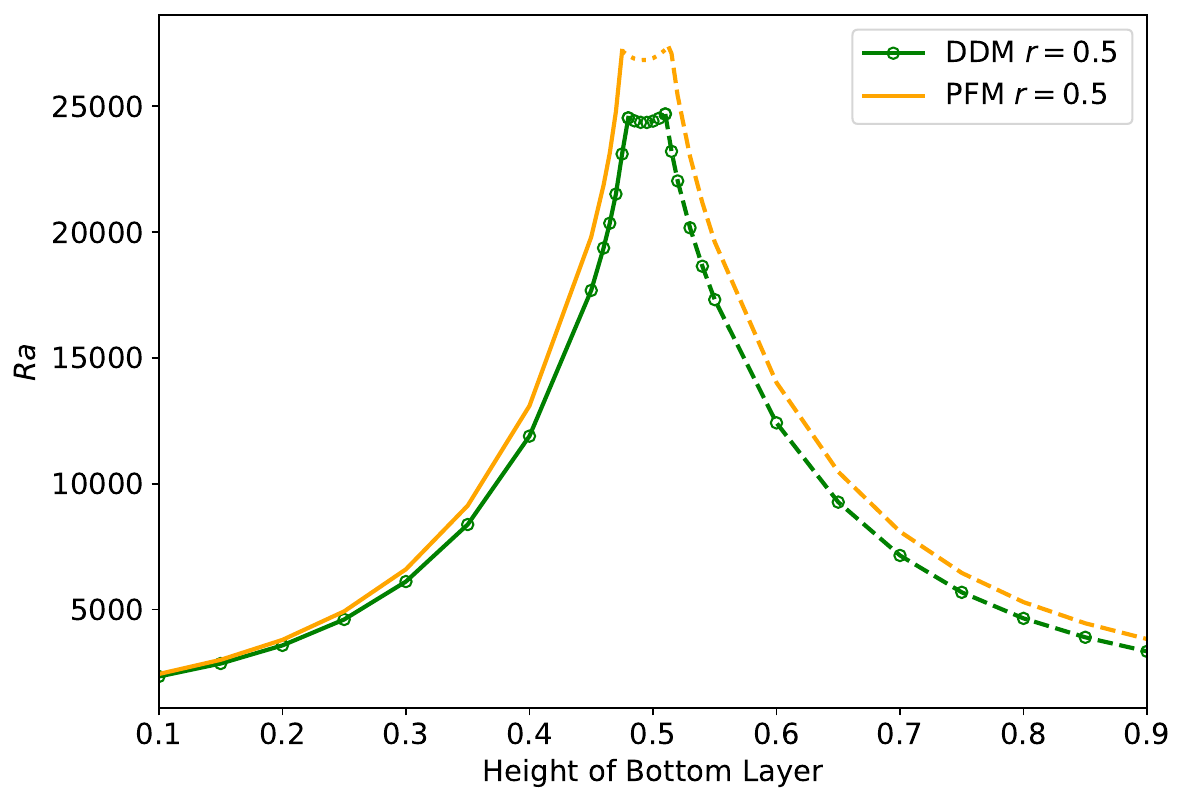}
    \caption{A comparison between the marginal stability curves obtained from domain decomposition method (DDM) and phase-field method (PFM) for sparingly miscible fluids.}
    \label{fig:RB_DDM_vs_PFM_r=0.5}
\end{figure}

\subsection{Onset of RB convection near UCST}

Adhering to our objective of characterising RB convection onset in diffuse interface systems, we now estimate the stability curves for different values of $a^*$  and $r$. For the given $\rho\beta\alpha\, (= 0.125)$ and $a^*$ values, we start from a purely immiscible state, i.e., $r=1$, and approach the UCST of the binary fluid combination by reducing $r$ to a value as low as $0.005$. The results of the former have already been presented in Fig.~\ref{fig:RB_neutral_curve_DDM_vs_PFM} wherein three distinct modes have been identified at different interfacial heights. These modes are marked as \enquote*{TOP}, \enquote*{BOT}, and \enquote*{OSC}, respectively. The part marked as \enquote*{TOP} corresponds to the \enquote*{upper dragging mode} wherein there is active buoyancy-driven convection in the top layer owing to its larger thickness. The bottom layer is passively driven by the continuity of velocity and shear stress at the interface. The \enquote*{BOT} curve represents the case vice-versa wherein the bottom layer is the driver, and the top layer is driven. At intermediate heights ranging between $0.485$ and $0.53$, the system manifests oscillatory (OSC) mode, wherein it transiently alters between a mechanically coupled state and a thermally coupled state. Here, both layers have equal propensities for primary excitation, and the system responds by oscillating between the two possible modes of coupling between the layers. In fact, on either side of this oscillatory range, the system exhibits stationary mechanical and thermal coupling modes depending on the properties chosen. The occurrence of oscillatory convection in a two-layer system is not always guaranteed. As mentioned earlier, a favourable combination of property ratios such as $\rho\beta\alpha$ being far from unity and $a^{*}$ being closer to unity is essential for the manifestation of oscillatory excitation.

Before performing the analysis for the other values of $r$, we shall briefly pause to ponder over the necessity of using the phase-field formulation for the current problem.
Figure~\ref{fig:RB_neutral_curve_DDM_vs_PFM} showed an excellent match of results between the present model and the sharp-interface approach, modelled via DDM. Taking a cue from this exact match, one might be tempted to use DDM in the binary fluid context wherein the layer properties are obtained from the equilibrium composition (Eq.~\eqref{eq:base_Phi}) at different temperatures. After all, the diffuse-interface thickness is much smaller than the layer heights for the range of $r$ considered in the present analysis. Fortunately, the answer to the above dilemma is evident from Fig.~\ref{fig:RB_DDM_vs_PFM_r=0.5}, where $r$ is $0.5$. The figure compares the phase-field model results with those of DDM, wherein the properties have been modified according to the equilibrium composition. The value of $r$ being $0.5$ implies that the fluid system is still far away from its consolute point, and yet, we see a noticeable difference between the neutral curves of PFM and DDM. The latter approach underpredicts the critical behaviour since the diffuse nature of the interface has been ignored. By correctly accounting for the dissipative effects in the small diffuse region, the PFM provides a more realistic picture of the onset behaviour. Thus, it becomes essential for any analysis involving multi-layer systems to first verify the value of $r$ before applying the sharp interface assumption.  

\begin{figure}
    \centering
    \includegraphics[scale=0.52]{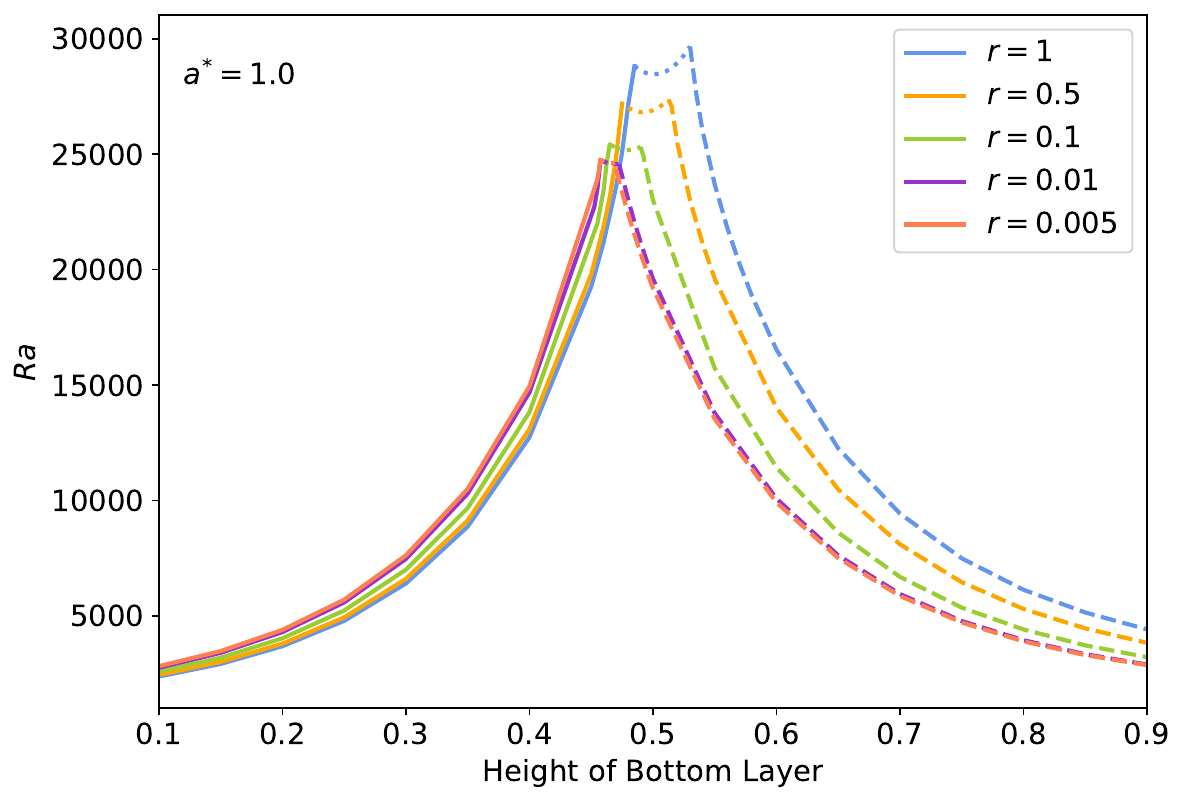}
    \caption{Neutral curves for different values of $r$ for $(\rho\beta\alpha)_r = 0.125, a^{*} = 1.0$}
    \label{fig:RB_Ra_vs_interface_poistion_var_r_$ρβκ_T=0.125$_astar=1.0}
\end{figure}

\begin{figure}
\centering
\includegraphics[scale=0.52]{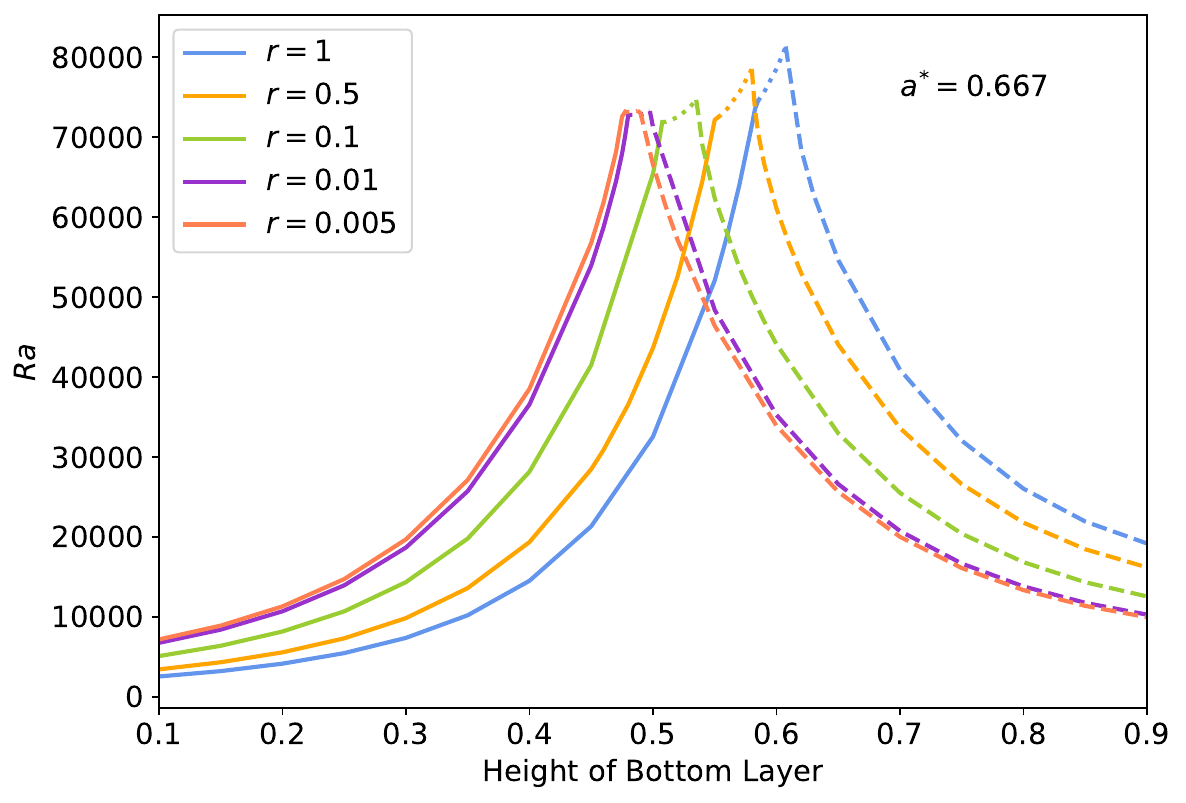} 
\label{fig:RB_Ra_vs_interface_poistion_var_r_$ρβκ_{T}=0.125$_astar=0.667_1}
\caption{Neutral curves for different values of $r$ for $(\rho\beta\alpha) = 0.125, a^{*} = 0.667$}
\label{fig:RB_Ra_vs_interface_poistion_var_r_$ρβκ_T=0.125$_astar=0.667}
\end{figure}

\begin{figure}
\centering
\includegraphics[scale=0.52]{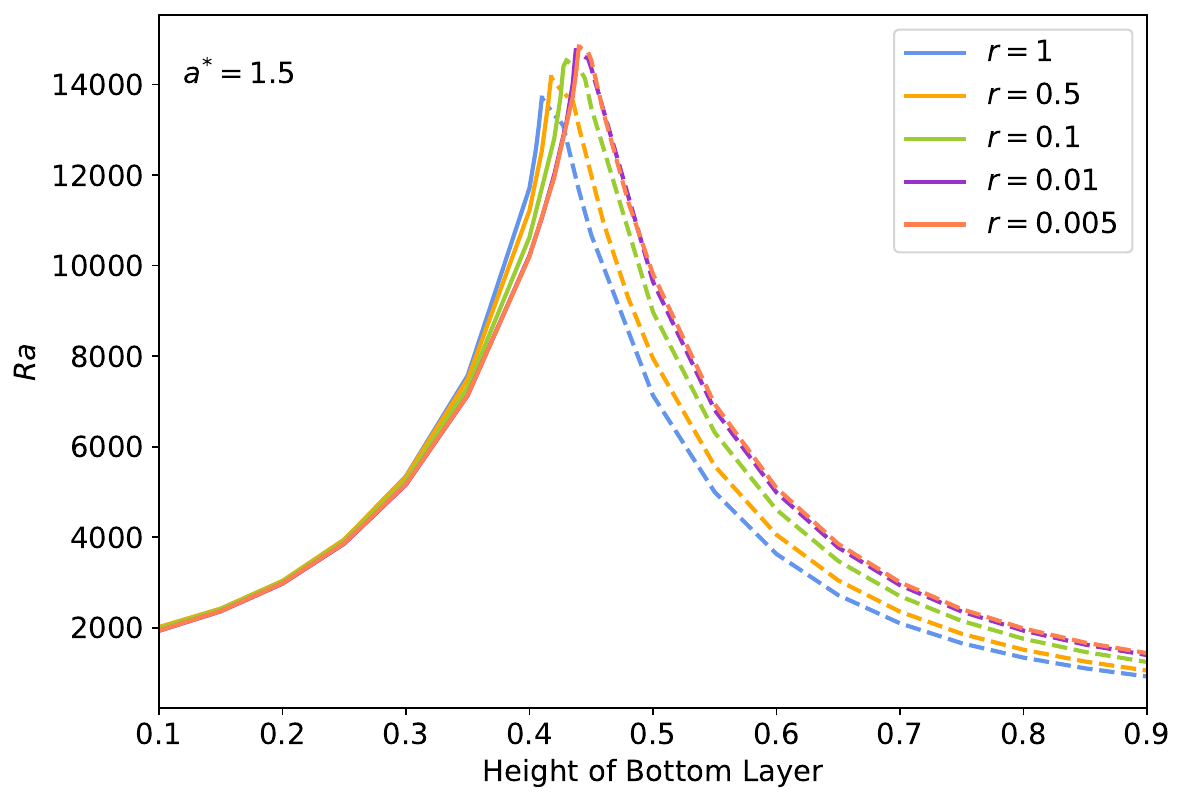} 
\label{fig:RB_Ra_vs_interface_poistion_var_r_$ρβκ_T=0.125$_astar=1.5_1}
\caption{Neutral curves for different values of $r$ for $(\rho\beta\alpha) = 0.125, a^{*} = 1.5$}
\label{fig:RB_Ra_vs_interface_poistion_var_r_$ρβκ_T=0.125$_astar=1.5}
\end{figure}

The characteristics of RB convection onset near UCST are now analysed by reducing the value of $r$. Figure~\ref{fig:RB_Ra_vs_interface_poistion_var_r_$ρβκ_T=0.125$_astar=1.0} shows the variation of the critical Rayleigh number with interfacial height for different $r$ values. It is evident that with the increased miscibility of fluids, the peaks of the marginal stability curves shift leftward, and the window for oscillatory onset becomes narrow. This decrease can be essentially attributed to the change in the equilibrium composition of the two layers that increases the apparent $\Tilde{\rho}\Tilde{\beta}\Tilde{\alpha}$ value, i.e., it approaches one. Also evident from Fig.~\ref{fig:RB_Ra_vs_interface_poistion_var_r_$ρβκ_T=0.125$_astar=1.0} is that the change in equilibrium composition has different influences on the \enquote*{TOP} and \enquote*{BOT} modes. Recall that for the present fluid system with $\rho\beta\alpha=0.125$ and $a^{*}=1.0$, the values of the relevant property ratios in the immiscible limit are $\nu=0.5$, $\beta=0.125$, $\alpha=0.5$. Thus, an increase in the system's temperature alters these ratios such that the thermal expansivity coefficient of the bottom layer increases, and so do the kinematic viscosity and thermal diffusivity. However, the increase in the former coefficient is significant compared to the latter dissipative effects, making the bottom layer more vulnerable to perturbations. Hence, there is an effective reduction in the critical $Ra$ value for the \enquote*{BOT} mode. The scenario for the top layer is the opposite, and we observe a marginal stabilisation of the \enquote*{TOP} mode.

Figure~\ref{fig:RB_Ra_vs_interface_poistion_var_r_$ρβκ_T=0.125$_astar=0.667} shows the critical Ra plots for the second system of present consideration, i.e., the one with $\rho\beta\alpha=0.125$ and $a^{*}=0.667$. Here, we see a behaviour similar to the one observed in the previous case. The peaks of the marginal curves swift leftward with the increase in base temperature, i.e. with reducing $r$. Note that for the present system, in the immiscible limit, the kinematic viscosity ratio is $2.53$, and any change in composition brought by the (base) temperature increase makes the bottom layer less dissipative. This, coupled with the increased thermal expansivity in the bottom layer, significantly reduces critical $Ra$ for the \enquote*{BOT} mode as it becomes easy to establish convection there. The change in composition also makes the increase in the critical $Ra$ for the \enquote*{TOP} mode more prominent compared to the $a^{*}=1.0$ system. Interestingly, the peak values of critical $Ra$ do not decrease here as much as in the previous case, mainly owing to the choice of fluid properties. Nonetheless, we observe a similar reduction in the window of the oscillatory convection, and the peak value settles around the interfacial height of $0.46-0.47$.

Lastly, we analyse the behaviour of the two-layer system with property ratio combinations, $\rho\beta\alpha=0.125$ and $a^{*}=1.5$. Note that the kinematic viscosity ratio of the system, $\nu$, is approximately 0.09875. Figure~\ref{fig:RB_Ra_vs_interface_poistion_var_r_$ρβκ_T=0.125$_astar=1.5} shows the marginal stability curves for different values of $r$. Interestingly, the curves undergo a rightward shift, though we observe a similar decrease in the window for oscillatory excitation in the system. The latter behaviour again relates to the increase in the apparent $\Tilde{\rho}\Tilde{\beta}\Tilde{\alpha}$ value that makes the occurrence of oscillatory excitation less probable. The rightward shift, however, occurs due to the large disparity in the kinematic viscosities of the two layers. The change in equilibrium composition due to the base temperature increases the viscosity of the bottom layer significantly, overwhelming the influence brought in by the increase in the thermal expansion coefficient. Consequently, the \enquote*{BOT} mode becomes more stable, requiring a larger temperature gradient to provoke instability in the system. The composition change reveals an opposite behaviour for the \enquote*{TOP} mode. 

From the analyses of the above three systems, it is evident that the propensity of a system to exhibit oscillatory convection decreases as it approaches UCST. The equilibrium composition at these states would have a lesser disparity in the property ratios. Notably, the combination $\Tilde{\rho}\Tilde{\beta}\Tilde{\alpha}$ approaches unity, and as shown by \cite{Renardy1996}, such systems would become devoid of any oscillatory convection. Of course, each system comes with its own drift pattern in the stability curves, essentially determined by the thermo-physical/transport properties of the fluids.



\section{Combined role of buoyancy and thermocapillarity}

Following the characterization of RB convection, we now analyze the influence of added thermocapillarity on the onset of convection in binary fluid systems. In this regard, we revert to the complete set of perturbation equations, i.e., from Eq.~\ref{eq:disc_ch} to Eq.~\ref{eq:disc_temp}, along with the associated boundary conditions mentioned in Eq.~\ref{eq:bouncon}. Despite the inclusion of Marangoni effects, the GEP still quantifies the criticality in terms of the Rayleigh number. This is on account of $\rm Ra$ and $\rm Ma$ being linked with each other through the additional input parameter, $\zeta\;(= \sigma_\theta/g\rho_2\beta_2 H^2)$. Consequently, the fluid combinations for the present analysis are characterized by four property groups such as a) the $\rho\beta\alpha$ combination, which is still maintained at 0.125, b) the critical height ratio, $a^*$, c) the miscibility transition parameter, $r$, that is varied between 1.0 and 0.005, and d) $\zeta$, which ranges between zero and 0.1. Before characterizing the influence of the new parameter, $\zeta$, we will briefly check the consistency of the Marangoni formulation.
   
\subsection{Consistency check}
\begin{figure}
\centering
\includegraphics[scale=0.52]{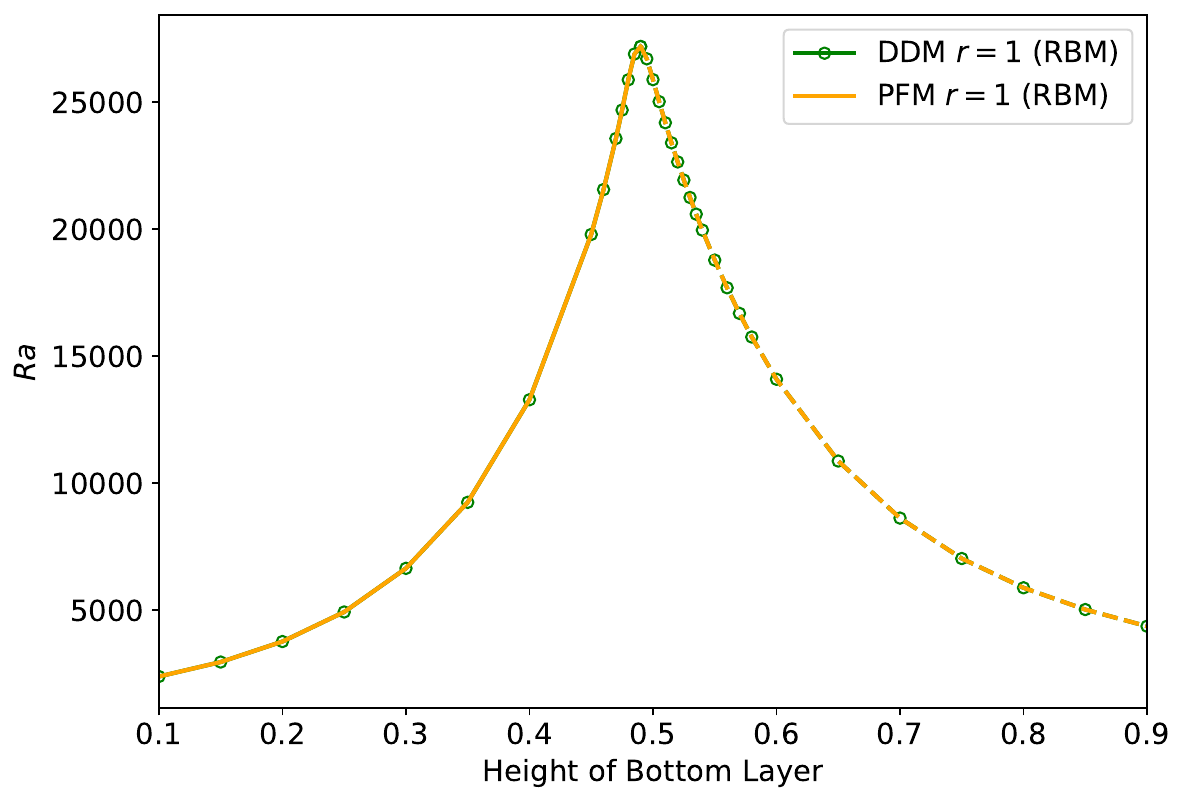}
\caption{A comparison between the curves of marginal stability for the RBM convection obtained from the domain decomposition method (DDM) and the phase-field method (PFM).}
\label{fig:RBM_DDM_vs_PFM}
\end{figure}

Repeating the analysis performed for the RB case, we now check the ability of the current diffuse RBM formulation to mimic the sharp-interface behaviour in the immiscible limit. The comparison is once again carried out against the DDM implementation of \cite{Diwakar2014JFM}. The property combinations associated with the fluid system are $\rho\beta\alpha = 0.125$, $r = 1.0$, $a^* = 1.0$, and $\zeta=0.01$. The value of $\epsilon/H$ is specified as $2 \times 10^{-4}$ to mimic the sharp interface behaviour.  Figure \ref{fig:RBM_DDM_vs_PFM} shows an exact match between the results of the current phase-field model in the sharp interface limit and the DDM implementation. Interestingly, the manifestation of oscillatory onset, as observed for the intermediate height range in Figure~\ref{fig:RB_neutral_curve_DDM_vs_PFM}, vanishes with the inclusion of Marangoni effect. The reason for this behaviour is discussed in the ensuing subsection.


\subsection{RBM convection in the immiscible limit}
\begin{figure}
\centering
    \includegraphics[scale=0.52]{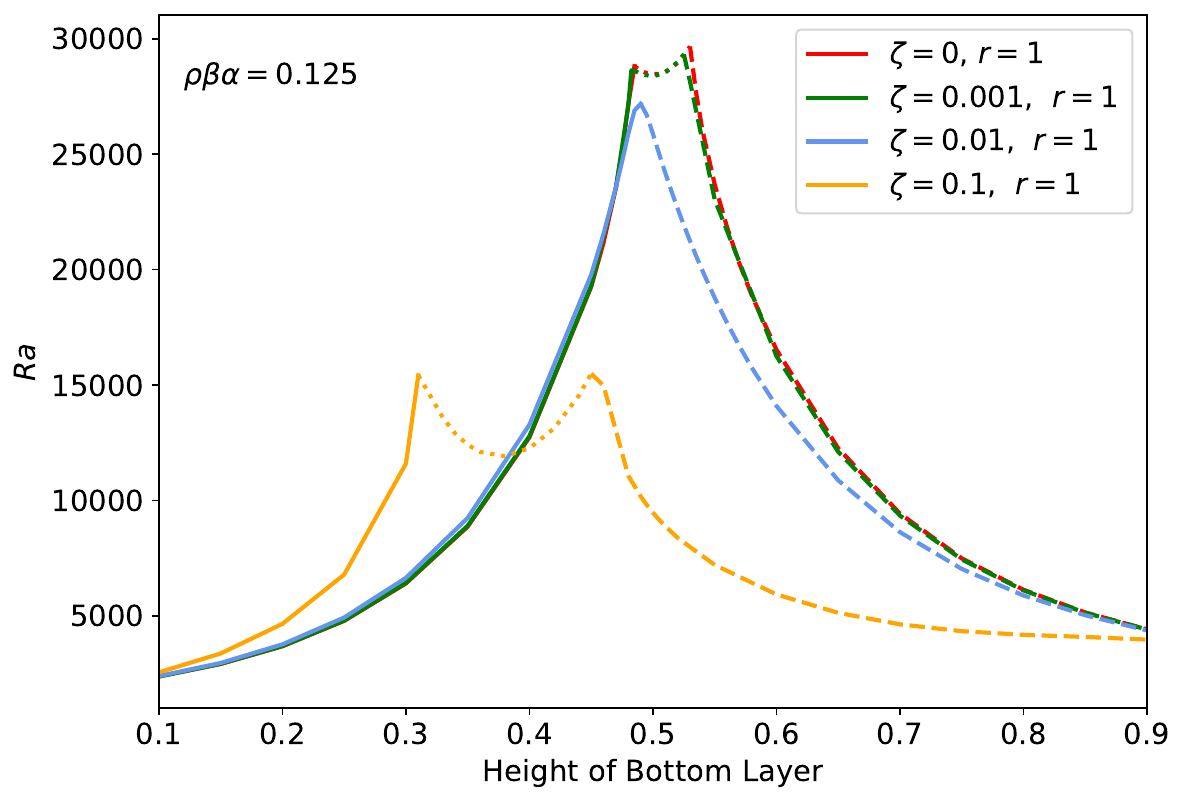}
    \caption{A comparison between neutral curves between RB and RBM convections. The magnitude of $\zeta$ corresponds to the strength of the surface tension gradients.}
    \label{fig:RB_and_RBM_marginal_stability_curves}
\end{figure}

Apart from the favourability of property ratios like $\rho\beta\alpha$ and $a^*$, the Marangoni effect also plays an important role in the occurrence of oscillatory convection in immiscible two-layer systems. \cite{Nepomnyashchy2004} demonstrated this influence using the silicone oil-water combination, wherein they showed that the experiment of \cite{Degen1998} would not have exhibited oscillatory convection if not for the presence of thermo-capillary effects. 

In a typical two-layer system, hot spots are formed along the fluid interface when the convection is dominant in the bottom layer. Since the Marangoni effect drives fluid away from a hot spot, it aids the underlying buoyancy-driven convection. The system thus becomes easily destabilised, resulting in lower criticality. The scenario is exactly the opposite for the top layer dominance, and the corresponding critical values are higher than those in the pure buoyancy-driven case where $\rm Ma = 0$. These diverse roles of the Marangoni effect, i.e., aiding convection sometimes and inhibiting it at other times, have their signature on the oscillatory onset of convection in the system. To understand them, we now gradually increase the magnitude of the thermo-capillary effect by increasing the $\zeta$ value of a fluid combination with property ratios $\rho\beta\alpha=0.125$, $a^{*}=1.0$, and $r=1$. Figure~\ref{fig:RB_and_RBM_marginal_stability_curves} shows the neutral curves obtained from the analysis, and it is interesting to note that the thermo-capillarity does not always increase the parametric window for oscillatory onset. Evidently, the window for oscillatory convection shrinks with the increase of $\zeta$ from zero and vanishes at around $\zeta=0.01$. With a further increase in the $\zeta$ value, oscillatory onset reappears. The reasons for this behaviour can be understood by extending the analysis of \cite{Renardy1996} to include the Marangoni effect. The details of the steps involved are shown in Appendix \ref{appB}, and for the sake of brevity, only the final equation is written below.

\begin{eqnarray}
    \lambda^{*} \Bigg[ \int_0^I \kappa|\Hat{\theta}_1|^2 \mathrm{dy} + \int_I^1 \alpha |\Hat{\theta}_2|^2 \mathrm{dy} \Bigg] & =
    &- \int_0^I \kappa \left( k^2 |\Hat{\theta}_1|^2 + \left| \frac{\mathrm{d}\Hat{\theta}_1}{\mathrm{dy}} \right|^2 \right) \mathrm{dy} - \int_I^1 \left( k^2 |\Hat{\theta}_2|^2 + \left| \frac{\mathrm{d}\Hat{\theta}_2}{\mathrm{dy}} \right|^2 \right) \mathrm{dy} \nonumber \\
    &&- \frac{\lambda}{Pr_2} \frac{\Lambda_2}{k^2 Ra} \Bigg[ \int_0^I \left( k^2 |\Hat{v}_1|^2 + \left| \frac{\mathrm{d}\Hat{v}_1}{\mathrm{dy}} \right|^2 \right) \mathrm{dy} \nonumber \\
    &&+ \int_I^1 \beta \alpha \left( k^2 \left|\Hat{v}_2\right|^2 + \left| \frac{\mathrm{d}\Hat{v}_2} {\mathrm{dy}} \right|^2 \right) {\mathrm{dy}} \Bigg] \nonumber \\
    &&- \frac{\Lambda_2}{k^2 Ra} \bigg[ \int_0^I \left| \frac{\mathrm{d}^2\Hat{v}_1}{\mathrm{dy}^2} \right|^2 \mathrm{dy} + \int_I^1 \frac{(\rho\beta\alpha)}{\eta}\left| \frac{\mathrm{d}^2\Hat{v}_2}{\mathrm{dy}^2} \right|^2 \mathrm{dy} \bigg] \nonumber \\
    &&+ \frac{\Lambda_2}{k^2 Ra} \left( \rho\beta\alpha - 1 \right) \frac{\mathrm{d}{\Hat{v}^*}_1}{\mathrm{dy}}  \frac{\mathrm{d}^2\Hat{v}_1}{\mathrm{dy}^2}\bigg|_{y=I} \nonumber \\
    &&+ \Lambda_2 (\rho\beta\alpha) \zeta \theta_1 \frac{\mathrm{d}{\Hat{v}^*}_1}{\mathrm{dy}}\bigg|_{y=I}
    \label{single_governig_equation_RBM}
\end{eqnarray}

The last two terms in the above equation, which are associated with the velocity derivatives at the interface, contribute to the non-self-adjointness of the system. Interestingly, the term involving the Marangoni effect competes with the term containing $\left(\rho\beta\alpha -1\right)$ value since the latter is negative. When the value of $\zeta$ is zero, the system would be non-self-adjoint simply by virtue of the favourable $\rho\beta\alpha$ (= 0.125 in the present case). For some non-trivial $\zeta$, the thermo-capillarity nullifies the $\rho\beta\alpha$ influence and makes the system self-adjoint. This results in the manifestation of stationary onset at all interfacial heights, as observed in Fig~\ref{fig:RBM_DDM_vs_PFM}. In the case of the present fluid combination (Fig~\ref{fig:RB_and_RBM_marginal_stability_curves}), the two influences cancel each other around $\zeta=0.01$. When $\zeta$ becomes larger than this value, the thermo-capillarity gains dominance; the system once again becomes non-self-adjointness and exhibits oscillatory convection.

\subsection{RBM convection in binary fluid system}

\begin{figure}
    \centering
    \includegraphics[scale=0.52]{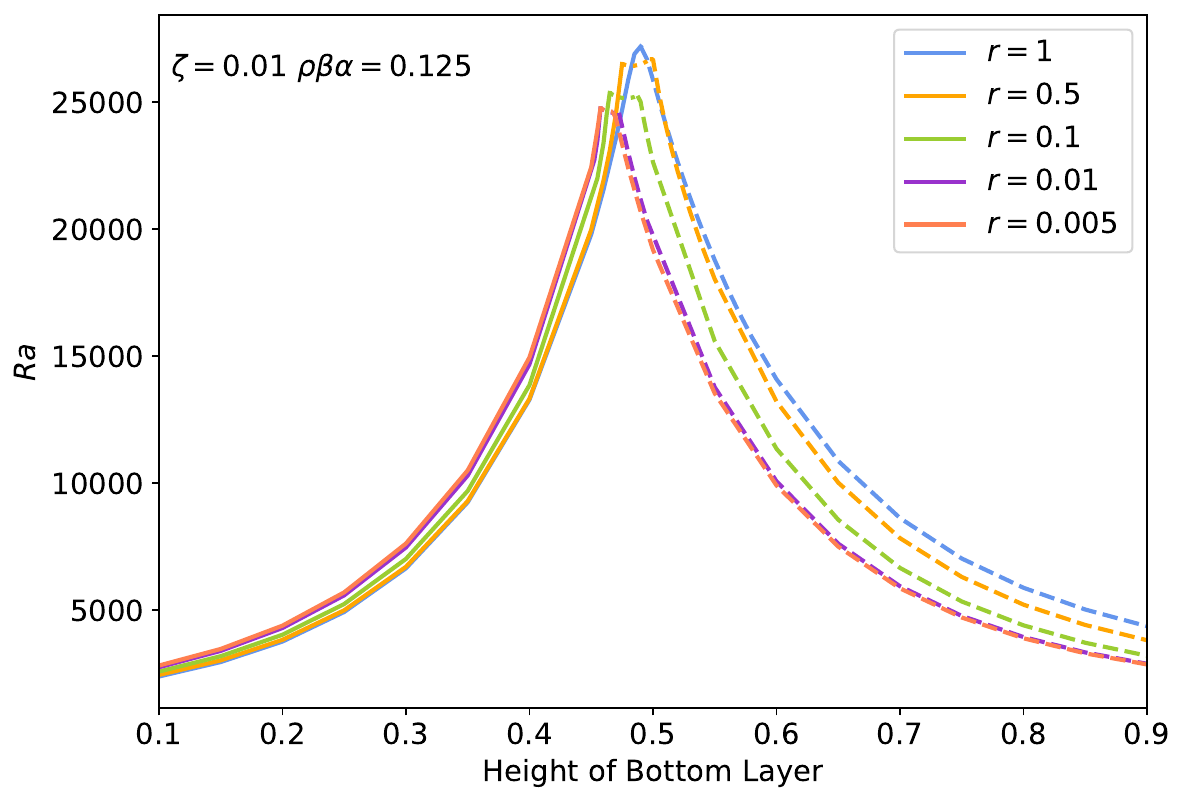}
    \caption{Marginal stability curves for different values of $r$ for $\zeta=0.01$, $(\rho\beta\alpha)=0.125$. A decrease in the value of $r$ represents more soluble mixtures.}
    \label{fig:RBM_Ra_vs_interface_poistion_var_r_Ups=0.01_RP=0.125.}
\end{figure}

\begin{figure}
    \centering
    \includegraphics[scale=0.52]{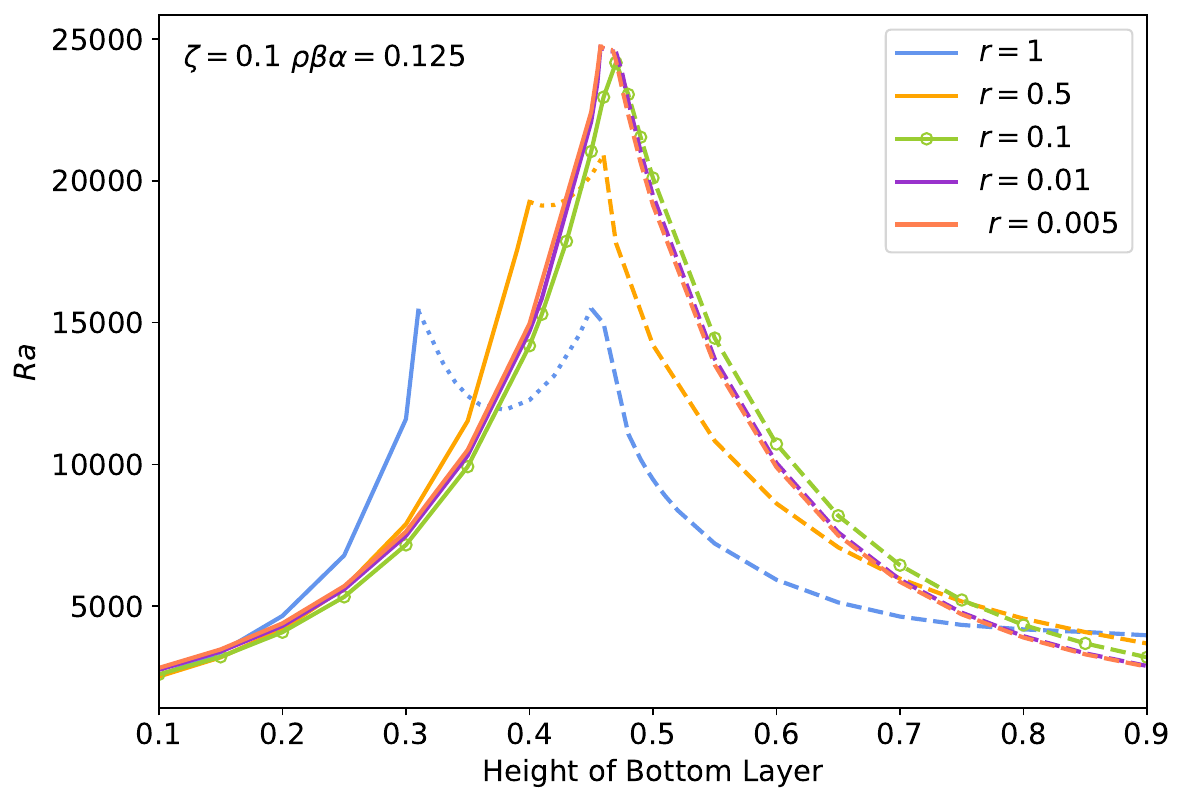}
\caption{Marginal stability curves for different values of $r$ for $\zeta=0.1$, $(\rho\beta\alpha)=0.125$. A decrease in the value of $r$ represents more soluble mixtures.}
\label{fig:RBM_Ra_vs_interface_poistion_var_r_Ups=0.1_RP=0.125.}
\end{figure}
\begin{figure}
    \centering
    \includegraphics[scale=0.52]{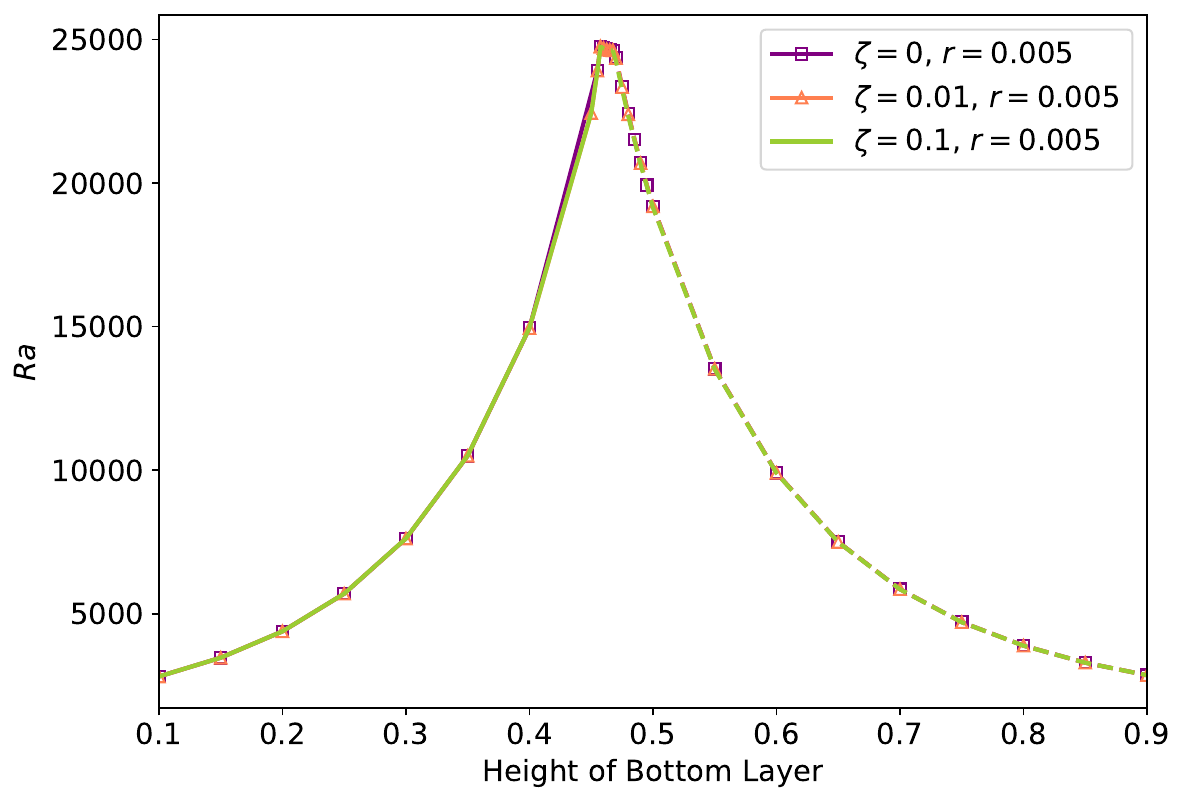}
    \caption{A comparison between the marginal stability curves between the RB and RBM convections closer to the critical temperature.}
    \label{fig:Ra_vs_interface_poistion_RB_vs_RBM_r=0.005}
\end{figure}

Moving away from the immiscible consideration, we will now discuss the onset characteristics of RBM convection in binary fluids. Understandably, both the coefficient of surface tension and its gradient w.r.t. temperature reduce to zero when the system's base temperature approaches UCST ($r\to0$). Despite this behaviour, the Marangoni effect still influences the flow onset pattern in the system, at least for moderate values of $r$. In order to understand these influences, the Marangoni effect is modelled presently by considering the mixing energy to be a linear function of temperature, i.e., $\Lambda = \Lambda_0 + \left(\partial\Lambda / \partial\theta\right) \theta$.  The decrease in the magnitude of the above mixing energy with the increase in interfacial thickness, i.e., with reducing $r$, suitably models the dwindling behaviour of the Marangoni effect near UCST. 

It is worth recalling that in the case of RB convection, the window of oscillatory instability invariably shrunk with the decrease in $r$ value. With the inclusion of the Marangoni effect, we will now see a completely different manifestation. Figures ~\ref{fig:RBM_Ra_vs_interface_poistion_var_r_Ups=0.01_RP=0.125.} and~\ref{fig:RBM_Ra_vs_interface_poistion_var_r_Ups=0.1_RP=0.125.} show the neutral curves for different values of $r$ at two constant $\zeta$ values. Note that the first system (Fig.~\ref{fig:RBM_Ra_vs_interface_poistion_var_r_Ups=0.01_RP=0.125.}) with $\rho\beta\alpha=0.125$, $a^{*}=1.0$, and $\zeta=0.01$ has no oscillatory transition in the pure immiscible state ($r=1$). With the decrease in $r$, the window for oscillatory onset interestingly increases until a point. However, closer to UCST, i.e., for very small values of $r$, the oscillatory window narrows as the Marangoni effect becomes feeble. In fact, in this region, the system exhibits behaviour similar to those observed for RB convection.

The second system (Fig.~\ref{fig:RBM_Ra_vs_interface_poistion_var_r_Ups=0.1_RP=0.125.}) with $\rho\beta\alpha=0.125$, $a^{*}=1.0$, and $\zeta=0.1$ possesses a relatively large window of oscillatory onset in the immiscible limit. The oscillatory window initially shrinks with the decrease in $r$. In fact, it completely vanishes at $r=0.1$. However, with the further reduction in $r$, the oscillatory window reappears and has features similar to the previous case.

Interestingly, the above two systems with $\zeta=0.01$ and $\zeta=0.1$ have manifested different behaviours. In the first case, we see that the window for oscillatory convection first increases and then decreases with the reduction of $r$. The situation is exactly the opposite in the second case. The reason for this divergent behaviour can be understood by revisiting Eq.~\eqref{single_governig_equation_RBM}. As mentioned in the previous subsection, the onset of oscillatory convection is governed by two interfacial terms containing $(\rho\beta\alpha-1)$  and $\zeta$, respectively. With the increase of interfacial thickness, i.e. with reducing $r$, the influences of both these terms dwindle. However, the Maragoni effect weakens much faster and plays no role for small $r$ values. This is evident from Fig. \ref{fig:Ra_vs_interface_poistion_RB_vs_RBM_r=0.005} wherein the neutral curves for different values of $\zeta$ are identical with the RB case. For moderate $r$ values, nevertheless, the terms compete to manifest diverse behaviour. In the first case with $\zeta=0.01$, the $(\rho\beta\alpha-1)$ term and the $\zeta$ term cancel each other at $r=1$. With the reduction in $r$, the increased solubility of fluids in each other reduces the disparities in the layer properties. As a result, $\Tilde{\rho}\Tilde{\beta}\Tilde{\alpha}$ approaches one, and the impact of the corresponding term weakens. At the same time, the influence of the $\zeta$ term increases as it is also multiplied by  $\rho\beta\alpha$. This leads to the occurrence of oscillatory onset for lower $r$ values. Eventually, this oscillatory window narrows as the interfacial thickness increases and the mixing energy density reduces. 



\begin{figure}
    \centering
    \includegraphics[scale=0.52]{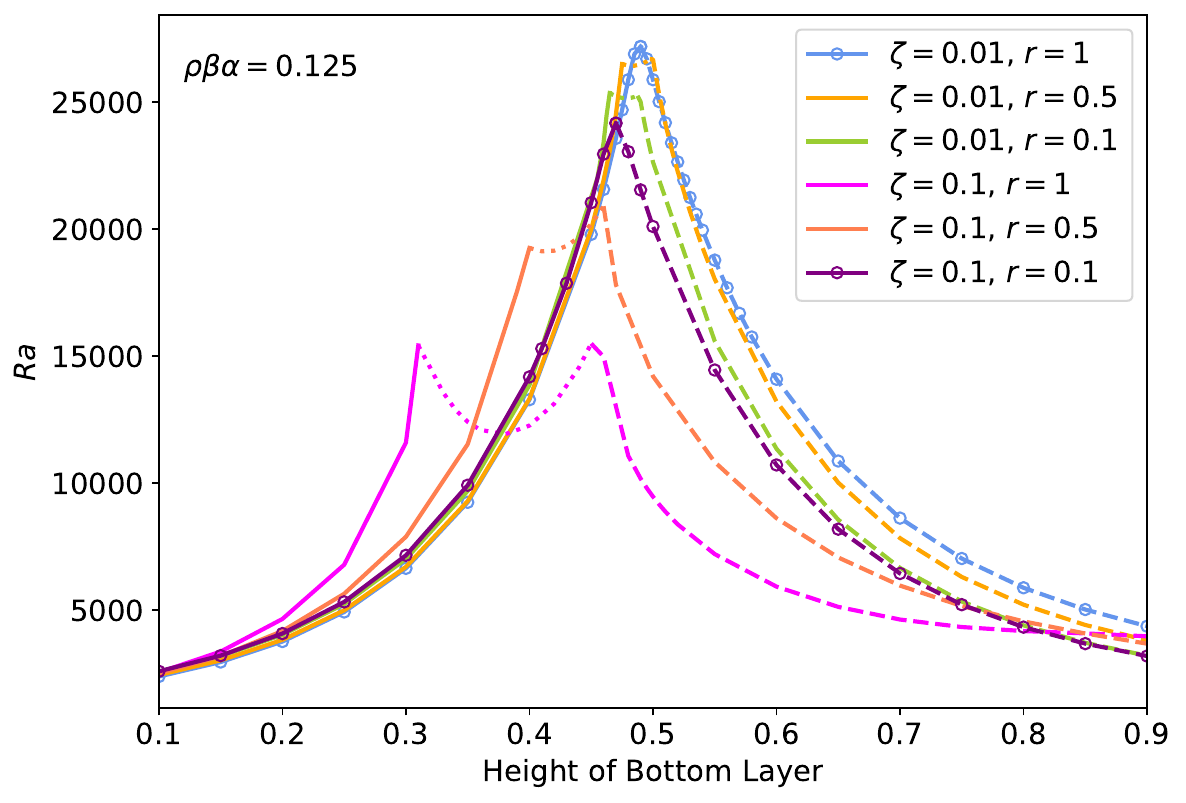}
    \caption{Marginal stability curves for different values of $\zeta$ and $r$ for $(\rho\beta\alpha)=0.125$. Markers (o) over the lines correspond to the non-oscillatory onset behaviour.}
    \label{fig:RBM_Ra_vs_interface_poistion_var_upsilon_RP=0.125}
\end{figure}

We can observe from the above discussions that the addition of the thermo-capillary effect can affect the onset characteristics in both ways. It can either inhibit or expand the oscillatory regime. In order to put things under proper perspective, we now take a combination of data as shown in Fig~\ref{fig:RBM_Ra_vs_interface_poistion_var_upsilon_RP=0.125}. Here, the system with \enquote*{$\zeta=0.01$} has no oscillatory onset in the immiscible limit, whereas it manifests a sizable window with the reduction of \enquote*{r}. On the other hand, the system with \enquote*{$\zeta=0.1$} exhibits oscillatory onset in the immiscible limit and becomes completely non-oscillatory at $r=0.1$. With further reduction of \enquote*{$r$}, it regains the zones of oscillatory excitation.

\begin{figure}
\centering
\includegraphics[scale=0.52]{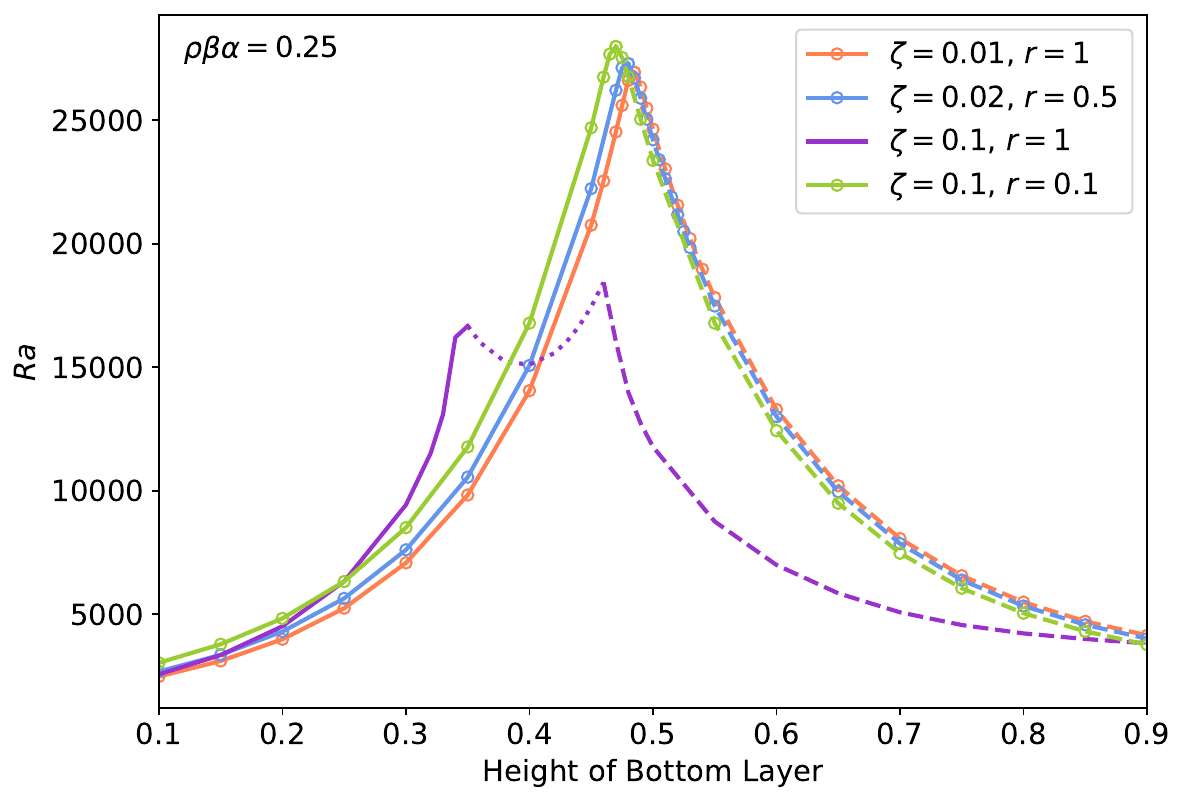}
\label{RBM_Ra_vs_interface_poistion_var_upsilon_RP=0.25}
\includegraphics[scale=0.52]{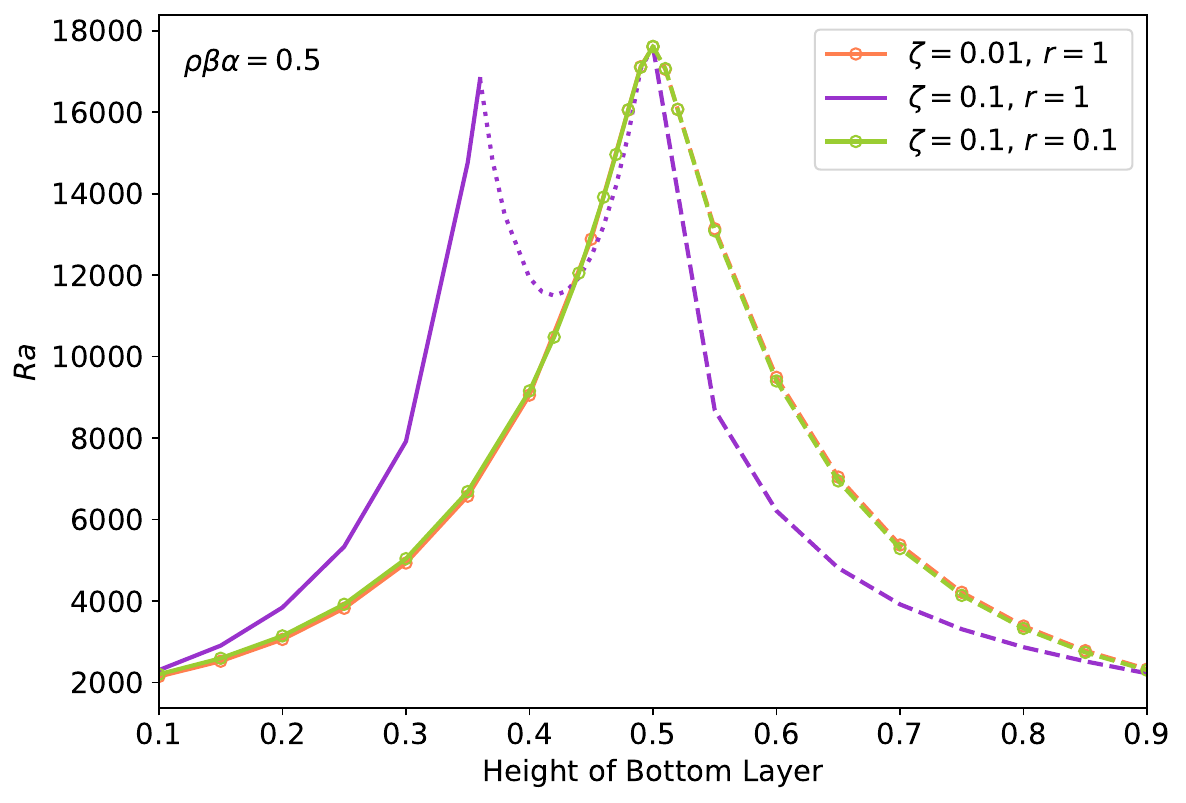}
\label{RBM_Ra_vs_interface_poistion_var_upsilon_RP=0.5}
\caption{Marginal stability curves for different values of $\zeta$ and $r$ for $(\rho\beta\alpha) = 0.25, 0.5$. Markers (o) over the lines correspond to the non-oscillatory onset behaviour.}
\label{fig:RBM_Ra_vs_interface_poistion_var_upsilon_RP=0.25,0.5.}
\end{figure}
It is worth remembering from the previous section that the nullification of the oscillatory regime in the completely immiscible limit occurs at $\zeta=0.01$. However, it is premature to state that the $r\zeta$ plays the same role as $\zeta$ in the pure immiscible state ($r=1.0$) without developing an expression similar to Eq~(\ref{single_governig_equation_RBM}) for the diffuse interface consideration. Nevertheless, we have been able to verify this claim for other systems with $(\rho\beta\alpha) = $ 0.25 and 0.5 as shown in Fig~\ref{fig:RBM_Ra_vs_interface_poistion_var_upsilon_RP=0.25,0.5.}. One can observe that the onset nature of the system becomes completely time-independent when $r\zeta \approx 10^{-2}$.


\section{Conclusion}
The analysis of the onset of RBM convection in binary fluids closer to UCST via a diffuse-interface approach reveals interesting features. Firstly, it is evident that the phase-field approach is potent enough to reveal the sharp-interface features in the truly immiscible limit exactly. Secondly, the analysis reveals that the effect of the actual diffused nature of the interface is too significant and cannot be ignored in the vicinity of the critical point. Thirdly, the propensity of the system to exhibit oscillatory convection decreases as it approaches UCST. The equilibrium composition at these states would have a lesser disparity in the property ratios. Particularly, the combination $\Tilde{\rho}\Tilde{\beta}\Tilde{\alpha}$ approaches unity, and as shown by \citet{Renardy1996}, such systems would become devoid of any oscillatory convection. Of course, each system would come with its own drift pattern in the stability curves, essentially determined by the thermo-physical/transport properties of the fluids. The current analysis of a two-layer RBM convection with a diffuse interface model provides some interesting information. In the pure immiscible limit, the Marangoni effect plays the dual role of suppressing the oscillatory convection in some range and enhancing it elsewhere. This is essentially due to the contribution played by two boundary terms, one involving $\left(\rho\beta\alpha - 1\right)$ and the other involving the non-dimensional surface tension gradient ($\zeta$). In some parametric ranges, these terms cancel each other, resulting in a self-adjoint system. In contrast, they do not nullify each other in other ranges to make the system non-self-adjoint and, thus, exhibit the possibility of oscillatory excitation. The role of solubility is quite intriguing as it acts in unison with the interfacial tension term to make the system either self-adjoint or non-self-adjoint. Thus, it comes with its own dual role, a behaviour not particularly evident for the RB instability in the immiscible limit. 

\section{Declaration of Interests}
The authors report no conflict of interest.

\appendix

\section{Recalculating spatial derivatives}\label{appA}

\begin{equation}
    \begin{split}
    \frac{\mathrm{d} \Xi}{\mathrm{d}\Tilde{y}} &= \frac{1}{g^{'}} \frac{\mathrm{d}\Xi}{\mathrm{d} y}\\
    \frac{\mathrm{d}^2 \Xi}{\mathrm{d}\Tilde{y}^2} &= \frac{1}{(g^{'})^2} \frac{\mathrm{d}^2 \Xi}{\mathrm{d}y^2} - \frac{g^{''}}{(g^{'})^3} \frac{\mathrm{d}\Xi}{\mathrm{d} y}\\
    \frac{\mathrm{d}^3 \Xi}{\mathrm{d}\Tilde{y}^3} &= \frac{1}{(g^{'})^3} \frac{\mathrm{d}^3 \Xi}{\mathrm{d} y^3} - 
    \frac{3g^{''}}{(g^{'})^4} \frac{\mathrm{d}^2 \Xi}{\mathrm{d} y^2} - 
    \frac{g^{'} g^{'''} - 3(g^{''})^2}{(g^{'})^5} \frac{\mathrm{d}\Xi}{\mathrm{d} y}\\
    \frac{\mathrm{d}^4\Xi}{\mathrm{d}\Tilde{y}^4} &= \frac{1}{(g^{'})^4} \frac{\mathrm{d}^4 \Xi}{\mathrm{d} y^4} - 
    \frac{6g^{''}}{(g^{'})^5} \frac{\mathrm{d}^3 \Xi}{\mathrm{d} y^3} - 
    \frac{4g^{'} g^{'''} - 15(g^{''})^2}{(g^{'})^6} \frac{\mathrm{d}^2 \Xi}{\mathrm{d} y^2} \\
    & - \frac{(g^{'})^2 g^{''''} - 10g^{'}g^{''}g^{'''} + 15(g^{''})^3}{(g^{'})^7} \frac{\mathrm{d}\Xi}{\mathrm{d}y}\\
    \end{split}
\end{equation}

\section{Revisiting theory to predict oscillatory onset in RBM convection}  \label{appB}

Following the typical normal mode expansion, i.e., the perturbations being proportional to $\exp(\lambda t + \mathrm{i}kx)$ as mentioned in \eqref{eq:normal_mode_expansion_form}, we get the temporal derivatives as 
\begin{equation}
    \centering
    \lambda \Hat{\boldsymbol{u}}_1 = -Pr_2 \bnabla \Hat{p}_1 + Ra Pr_2 \Hat{\theta}_1 \boldsymbol{1}_y + Pr_2\nabla{^2} \Hat{\boldsymbol{u}}_1
\end{equation}
\begin{equation}
    \centering
    \lambda \Hat{\boldsymbol{u}}_2 = -\rho Pr_2 \bnabla \Hat{p}_2 + \frac{Ra Pr_2} \beta \Hat{\theta}_2 \boldsymbol{1}_y + \frac{Pr_2} {\nu}\nabla{^2}\Hat{\boldsymbol{u}}_2
\end{equation}
\begin{equation}
    \centering
    \lambda \Hat{\theta}_1 = - \Hat{v}_1\Lambda{_1} + \nabla{^2}\Hat{\theta}_1
\end{equation}
\begin{equation}
    \centering
    \lambda \Hat{\theta}_2 = - \Hat{v}_2\Lambda{_2} + \frac{1} {\kappa_T}\nabla{^2} \Hat{\theta}_2
\end{equation}

Operating with $\boldsymbol{1}_y \cdot \bnabla \times \bnabla$, the momentum equations take the following forms 
\begin{equation}
    \centering
    \lambda{\nabla{^2} \Hat{v}_1} = -k^2 Ra Pr_2 \Hat{\theta}_1 + Pr_2\nabla{^4} \Hat{v}_1, 
\end{equation}
\begin{equation}
    \centering
    \lambda{\nabla{^2} \Hat{v}_2} = -k^2 \frac{Ra Pr_2}{\beta} \Hat{\theta}_2 + \frac{Pr_2}{\nu} \nabla{^4} \Hat{v}_2.
\end{equation}

Eventually, the set of governing equations can be written as
\begin{equation}
    \centering
    \frac{\lambda}{Pr_2}\nabla{^2} \Hat{v}_1 = -k^2 Ra \Hat{\theta}_1 + \nabla{^4} \Hat{v}_1
    \label{momentum_equation_1}
\end{equation}
\begin{equation}
    \centering
    \frac{\lambda}{Pr_2}\nabla{^2} \Hat{v}_2 = -k^2 \frac{Ra}{\beta} \Hat{\theta}_2 + \frac{1}{\nu}\nabla{^4} \Hat{v}_2
    \label{momentum_equation_2}
\end{equation}
\begin{equation}
    \centering
    \lambda\Hat{\theta}_1 = - \Hat{v}_1 \Lambda_1 + \nabla{^2} \Hat{\theta}_1
    \label{energy_equation_1}
\end{equation}

\begin{equation}
    \centering
    \lambda \Hat{\theta}_2 = - \Hat{v}_2\Lambda{_2} + \frac{1} {\kappa_T}\nabla{^2}\Hat{\theta}_2
    \label{energy_equation_2}
\end{equation}

Equations~(\ref{energy_equation_1}) and~(\ref{energy_equation_2}) are multiplied by $\kappa {\Hat{\theta}^*}_1$ and $\alpha {\Hat{\theta}^*}_2$ respectively. The addition of complex conjugate of the resultant equations gives
\begin{equation}
\begin{split}
    \lambda^{*} \left[\int_0^I \kappa |\Hat{\theta}_1|^2 dy + \int_I^1 \alpha |\Hat{\theta}_2|^2 dy \right] 
    &= \int_0^I \kappa\Hat{\theta}_1\nabla^2{\Hat{\theta}^*}_1 dy + \int_I^1 \alpha \Hat{\theta}_2\nabla^2{\Hat{\theta}^*}_2 dy \\
    &- \int_0^I \Lambda_2 {\Hat{v}^*}_1\Hat{\theta}_1 - \int_I^1 \alpha \Lambda_2 {\Hat{v}^*}_2\Hat{\theta}_2.
\end{split}
\end{equation}

Equations~(\ref{momentum_equation_1}) and~(\ref{momentum_equation_2}) are multiplied by $\Lambda_2 {v^*}_1$ and $\alpha\Lambda_2{v^*}_2$ respectively to eliminate the cross products like  \(\Bar{v}_1\theta_1\) and  \(\Bar{v}_2\theta_2\) from the momentum and energy equations.

Thus, the final form of the modified governing equation is given as 

\begin{eqnarray}
    \lambda^{*} \Bigg[ \int_0^I \kappa|\Hat{\theta}_1|^2 \mathrm{dy} + \int_I^1 \alpha |\Hat{\theta}_2|^2 \mathrm{dy} \Bigg] & =
    &- \int_0^I \kappa \left( k^2 |\Hat{\theta}_1|^2 + \left| \frac{\mathrm{d}\Hat{\theta}_1}{\mathrm{dy}} \right|^2 \right) \mathrm{dy} - \int_I^1 \left( k^2 |\Hat{\theta}_2|^2 + \left| \frac{\mathrm{d}\Hat{\theta}_2}{\mathrm{dy}} \right|^2 \right) \mathrm{dy} \nonumber \\
    &&- \frac{\lambda}{Pr_2} \frac{\Lambda_2}{k^2 Ra} \Bigg[ \int_0^I \left( k^2 |\Hat{v}_1|^2 + \left| \frac{\mathrm{d}\Hat{v}_1}{\mathrm{dy}} \right|^2 \right) \mathrm{dy} \nonumber \\
    &&+ \int_I^1 \beta \alpha \left( k^2 \left|\Hat{v}_2\right|^2 + \left| \frac{\mathrm{d}\Hat{v}_2} {\mathrm{dy}} \right|^2 \right) {\mathrm{dy}} \Bigg] \nonumber \\
    &&- \frac{\Lambda_2}{k^2 Ra} \bigg[ \int_0^I \left| \frac{\mathrm{d}^2\Hat{v}_1}{\mathrm{dy}^2} \right|^2 \mathrm{dy} + \int_I^1 \frac{(\rho\beta\alpha)}{\eta}\left| \frac{\mathrm{d}^2\Hat{v}_2}{\mathrm{dy}^2} \right|^2 \mathrm{dy} \bigg] \nonumber \\
    &&+ \frac{\Lambda_2}{k^2 Ra} \left( \rho\beta\alpha - 1 \right) \frac{\mathrm{d}{\Hat{v}^*}_1}{\mathrm{dy}}  \frac{\mathrm{d}^2\Hat{v}_1}{\mathrm{dy}^2}\bigg|_{y=I} \nonumber \\
    &&+ \Lambda_2 (\rho\beta\alpha) \zeta \Hat{\theta}_1 \frac{\mathrm{d}{\Hat{v}^*}_1}{\mathrm{dy}}\bigg|_{y=I}.
\end{eqnarray}

The boundary value terms arise from integration by parts with respect to $y$ with conditions: $\Hat{v}_1 = \Hat{v}_2 = 0$ (non-deformable interface), $\frac{\mathrm{d} \Hat{v}_1}{\mathrm{dy}} = \frac{\mathrm{d} \Hat{v}_2}{\mathrm{dy}}$ (derived from continuity of tangential velocity), $\frac{\mathrm{d}^2 \Hat{v}_1}{\mathrm{dy}^2} = \frac{1}{\eta}\frac{\mathrm{d}^2 \Hat{v}_2}{\mathrm{dy}^2} + k^2 Ma \Hat{\theta}_1$ (continuity of shear stress), and $\kappa \frac{\mathrm{d} \Hat{\theta}_1}{\mathrm{dy}} = \frac{\mathrm{d} \Hat{\theta}_2}{\mathrm{dy}}$ (continuity of heat flux) at the interface.

\bibliographystyle{jfm}
\bibliography{jfm}

\end{document}